\documentclass[10pt,onecolumn]{article}

\usepackage[a4paper,margin=1in]{geometry}
\usepackage{authblk}                
\usepackage{graphicx}
\usepackage{physics}
\usepackage{color,comment}
\usepackage{amsmath,amssymb,amsfonts}
\usepackage{siunitx}                
\usepackage{physics}                
\usepackage[hidelinks]{hyperref}
\usepackage{booktabs}
\usepackage{microtype}
\usepackage[capitalise]{cleveref}
\usepackage[numbers,sort&compress]{natbib}
\usepackage[capitalise]{cleveref}
\usepackage{subcaption} 
\usepackage{xspace}

\newcommand{\AU}{\ensuremath{\mathrm{AU}}\xspace}
\newcommand{\GeV}{\ensuremath{\mathrm{GeV}}\xspace}
\newcommand{\TeV}{\ensuremath{\mathrm{TeV}}\xspace}
\newcommand{\nT}{\ensuremath{\mathrm{nT}}\xspace}
\newcommand{\kmps}{\ensuremath{\mathrm{km\,s^{-1}}}\xspace}
\newcommand{\degs}{\ensuremath{^\circ}\xspace}

\newcommand{\Ecrit}{\ensuremath{E_{\mathrm{crit}}}\xspace}
\newcommand{\Br}{\ensuremath{B_{r}}\xspace}
\newcommand{\Bth}{\ensuremath{B_{\theta}}\xspace}
\newcommand{\Bphi}{\ensuremath{B_{\phi}}\xspace}
\newcommand{\Vsw}{\ensuremath{V_{\mathrm{sw}}}\xspace}

\title{\textbf{Through the Heliospheric Lens: Directional Deflection of High-Energy Cosmic-Ray Electrons and Positrons}}
\author[1,2]{Stefano Profumo}
\author[1,3]{Aria Koul}
\author[1,4]{Anika Malladi}
\author[1,5]{Benjamin Schmitt}
\affil[1]{\small Department of Physics, University of California, Santa Cruz, Santa Cruz, CA, 95064, USA}
\affil[2]{\small Santa Cruz Institute for Particle Physics, University of California, Santa Cruz, Santa Cruz, CA, 95064, USA}
\affil[3]{\small Department of Information, University of California, Berkeley, Berkeley, CA, 94720, USA}
\affil[4]{\small Department of Physics, City College of New York, New York, NY, 10031, USA}
\affil[5]{\small Department of Physics, California Polytechnic State University, San Luis Obispo, CA,  93407, USA}

\begin{document}
\maketitle

\begin{abstract}
\noindent
We investigate how the large–scale heliosphere alters the arrival directions of high–energy cosmic–ray electrons and positrons and ask if and when this ``heliospheric lens’’ can be ignored for anisotropy and source–association studies -- an especially timely topic given, for instance, the persistent cosmic-ray positron fraction and its unknown origin. Using a modular back–tracing framework, we explore a set of widely used magnetic–field descriptions—from a Parker spiral baseline to more structured configurations that include latitudinal wind contrasts, Smith–Bieber–type azimuthal strengthening, and tilted or wavy heliospheric current sheets. {\color{black}We model the deterministic deflections of high-energy cosmic-ray electrons and positrons (CREs) induced by large-scale heliospheric magnetic-field structures using a back-tracing approach.  
Our results apply to CREs above tens of GeV, where diffusion, convection, and adiabatic energy losses play a subdominant role; these processes are neglected in the present study and will be addressed in future work.} Across these models the picture is consistent: most bending is accumulated within the inner tens of astronomical units and decreases rapidly with energy. Field choices and solar–cycle geometry set the overall normalization, with stronger spiral winding or a more highly tilted current sheet producing larger deflections at the same energy. Differences between electrons and positrons are most apparent at lower energies, where drift histories and current–sheet encounters diverge, and become increasingly small at multi–TeV energies. We summarize these trends with a practical threshold energy describing when heliospheric bending falls below an instrument’s angular resolution, and we verify that our conclusions are robust to numerical settings. For current instruments, heliospheric effects can usually be treated as a small correction at the highest energies, while sub–TeV analyses benefit from a calibrated envelope that accounts for plausible field configurations during the observing epoch.

\end{abstract}



\section{Introduction}
\label{sec:intro}

High-energy (HE) cosmic-ray electrons and positrons (CREs; energies $\gtrsim$~few~\GeV) are powerful probes of recent particle acceleration in the local Galaxy because radiative cooling shortens their propagation horizon and look-back time. As instruments push toward sub-degree angular studies from hundreds of \GeV{} to multi-\TeV, even modest magnetic bending on the final leg to Earth can bias anisotropy measurements and source localization if not modeled. The large-scale heliospheric magnetic field (HMF) is the coronal magnetic field advected outward by the solar wind and wound into an Archimedean (Parker) spiral by solar rotation \citep{Parker1958}, and authoritative reviews summarize its multi-scale structure, solar-cycle variability, and consequences for charged-particle transport \citep{Owens2013,Potgieter2013}. In the Parker picture, $\Br\!\propto\!r^{-2}$ and $\Bphi\!\propto\!-(\Omega_\odot r/\Vsw)\Br$, but observed departures arise from time dependence, the heliospheric current sheet (HCS), shear in the solar wind, and latitudinal structure; classic transport theory established the central role of gradient/curvature drifts and HCS crossings for charge-sign–dependent modulation in spiral-like fields \citep{Jokipii1977}. 

Before turning to model comparisons, it is helpful to recall why latitudinal structure in the heliosphere is empirically well constrained. \emph{Ulysses} was a joint ESA–NASA deep–space mission launched in 1990 that, after a Jupiter gravity assist, entered a highly inclined, out–of–ecliptic orbit enabling the first in situ surveys over the Sun’s polar caps. Over three polar passes across two solar cycles, \emph{Ulysses} combined magnetometer and solar–wind plasma measurements to map how field strength, sector structure, and flow speed vary with heliographic latitude and cycle phase \citep{Balogh1995,McComas2000,SmithBalogh2001}. Empirically, \emph{Ulysses} mapped the latitudinal organization of the solar wind and magnetic field: during solar minimum, slow equatorial wind ($\sim$350--450~\kmps) contrasts with fast polar streams ($\sim$650--800~\kmps), while solar maximum yields a more complex state \citep{McComas2000,Smith2001HCS}. These flows interact with coronal topology to shape the HCS, whose tilt and waviness evolve with the cycle and imprint sector structure. Beyond the basic Parker winding, the azimuthal field is often enhanced relative to simple expectations—frequently parameterized via a Smith--Bieber–type prescription \citep{SmithBieber1991}—and nonradial footpoint motion can introduce a small but finite $\Bth$ (``Fisk-like'' fields) \citep{Fisk1996}. The coronal field that seeds the HMF is commonly modeled with potential-field source-surface (PFSS) extrapolations \citep{AltschulerNewkirk1969,Schatten1969}, which, despite idealizations, capture much of the global open/closed topology and are widely used for heliospheric boundary conditions and HCS morphology (with refinements such as Wang--Sheeley \citep{WangSheeley1992}) and the current-sheet source-surface (CSSS) model that improves realism by allowing explicit current sheets and non-spherical source surfaces \citep{ZhaoHoeksema1995}. These frameworks link photospheric magnetograms to large-scale topology (open polar fields, streamer belts) that governs HCS geometry and charged-particle access.

{\color{black}Several works have investigated cosmic-ray transport in MHD backgrounds to account for small- and medium-scale anisotropy features observed at TeV energies.  
For instance, \citet{Giacinti2012_AnisotropyTurbulence} demonstrated that pitch-angle scattering in a turbulent local interstellar magnetic field can produce the angular power spectrum measured by ground-based observatories, while \citet{Desiati2013_AnisotropyMHD} showed that global anisotropy patterns can arise from the interplay of ordered and turbulent components of the local field.  
More recent studies have used three-dimensional MHD simulations to follow charged-particle trajectories and reproduce the observed amplitude and angular structure of TeV-scale anisotropies \citep{Lazarian2013_CRTransport, Mertsch2015_SmallScaleAniso}.  
We now cite these works and briefly discuss in the Introduction how our approach differs in scope: unlike these studies, which address interstellar propagation on parsec scales, our analysis focuses on the deterministic heliospheric bending of high-energy leptons within tens of astronomical units from the Sun.}

Direct solar-system evidence for degree- and sub-degree-scale deflection comes from the Sun shadow: a deficit of multi-TeV cosmic rays at the apparent solar position whose depth and centroid vary over the solar cycle. Tibet AS$\gamma$, HAWC, and IceCube (among others) have measured this time-dependent shadow and used it to test heliospheric magnetic-field models \citep{Amenomori2000SunShadow,HAWC2015SunShadow,IceCube2021SunShadow}. {\color{black}Complementary observations by ARGO--YBJ have demonstrated that the displacement of the Sun shadow can be used to infer the mean interplanetary magnetic field carried by the solar wind \citep{Aielli2011_ARGOIMF}, while recent measurements by LHAASO--WCDA extend Sun-shadow mapping to higher energies and improved angular resolution \citep{Zhang2023_LHAASOSunShadow}.} Modeling with PFSS/CSSS reproduces many observed features and highlights the sensitivity of the deflection pattern to the coronal topology and solar-cycle phase \citep{Tjus2020}. Although most Sun-shadow studies concern hadronic cosmic rays at energies {\color{black} similar or greater than a TeV, they benchmark the magnitude and temporal modulation of directional effects at the degree scale} relevant also to high-energy leptons. {\color{black}On the lepton side, recent searches for large-scale anisotropies in cosmic-ray electrons and positrons by CALET and DAMPE \citep{Motz2017_CALETAniso,Munoz2019_DAMPEAniso}, together with Fermi--LAT and AMS--02 \citep{Ackermann2010,Abdollahi2017,AMSICRC2021}, have achieved sub-percent dipole sensitivities from tens of GeV to multi-TeV energies, with no significant detections to date.} Interpreting these limits---and any future signals---requires quantifying the extent to which heliospheric bending may dilute, reorient, or otherwise bias an upstream Galactic anisotropy.

For anisotropy and source-association studies, three practical questions naturally arise: (i) for a given instrument angular resolution $\theta_{\rm inst}$, above what energy does heliospheric bending become negligible; (ii) how do model choices (Parker vs.\ dipole+quadrupole+current-sheet (DQCS) vs.\ DQCS with spiral and Smith--Bieber) and solar-cycle scalars (\Br(1~\AU), $\Vsw$) shift that threshold; and (iii) how strongly do HCS geometry (tilt, waviness) and charge sign modulate the directional response across the sky? These questions are at the core, and motivate, the present study. Addressing them requires a relativistic, numerically controlled back-tracing framework that isolates the heliospheric leg, supports modular field components, and summarizes deflections statistically (mean and percentiles over random sky directions). It also benefits from reporting the threshold energy $\Ecrit(\theta_{\rm inst})$ at which the mean bending falls below a chosen angular budget.

{\color{black}A comprehensive description of cosmic-ray transport in the heliosphere would require the inclusion of spatial diffusion, convection with the solar wind, drift motions in the large-scale heliospheric magnetic field, and adiabatic energy losses.  
These effects dominate at energies below several tens of GeV and depend sensitively on solar-cycle conditions.  
Our present back-tracing analysis neglects such processes and therefore applies primarily to the regime where particle trajectories are quasi-deterministic and energy changes are negligible.}

In what follows we present a streamlined, reproducible framework to quantify direction stability for HE CREs in the heliosphere, spanning tens of \GeV{} to multi-\TeV{} energies, and we make the following specific contributions:
\begin{itemize}
  \item \textbf{Modular back-tracing framework.} We develop a relativistic Lorentz-code with fourth-order Runge--Kutta integration, speed renormalization, and mild timestep adaptation $\propto 1/|B|$, designed to isolate the heliospheric contribution to directional bending while maintaining numerical stability. The implementation exposes model and numerical controls (steps per gyroperiod, outer stop radius, seeds, and sampling) and outputs per-run provenance together with CSV data products to ensure reproducibility and straightforward cross-checks.
  \item \textbf{Magnetic-field models and geometry.} We implement (i) a Parker spiral baseline \citep{Parker1958}; (ii) a DQCS field capturing dipole+quadrupole structure with an equatorial current-sheet term \citep{Banaszkiewicz1998}; (iii) a DQCS+\emph{spiral} augmentation that adds Parker-like $\Bphi$ with a Smith--Bieber enhancement and latitude-dependent $\Vsw(\theta)$ \citep{SmithBieber1991}; and (iv) a Fisk-like perturbation with a controlled $\Bth$ admixture \citep{Fisk1996}. We also include a tilted, wavy HCS (tilt, amplitude, phase, wave number $m$), enabling systematic scans of sector morphology and its charge-sign–dependent consequences.
  \item \textbf{Deflection statistics vs.\ energy.} For each model and parameter choice we compute the mean angular deflection and dispersion (16--84\% band) over random sky directions as a function of energy, for electrons ($q=-e$) and positrons ($q=+e$). {\color{black}In the following, we define the \emph{mean angular deflection} $\langle\Delta\theta\rangle$ as the average angular separation between the initial and final directions of the back-traced particles, computed over an isotropic ensemble at 1 AU.  
The associated \emph{dispersion} $\sigma_{\Delta\theta}$ corresponds to the standard deviation of the same distribution, providing a quantitative measure of its angular spread.  
These definitions are purely geometric and independent of the particle flux normalization, thus serving as operative quantities characterizing the directional distortion induced by the heliospheric magnetic field.} We summarize results via a practically useful metric, $\Ecrit(\theta_{\rm inst})$, the energy where the mean deflection falls below a target angular budget set by an instrument’s point-spread function.
  \item \textbf{Sensitivity to solar-cycle scalars and geometry.} We quantify how realistic ranges in \Br(1~\AU) (e.g., 3--7~\nT), equatorial $\Vsw$ (350--500~\kmps) with fast-polar contrast (650--800~\kmps), and HCS tilt/waviness (0--30$^\circ$, $m=1,2$) shift deflection curves and $\Ecrit$. We isolate which parameters dominate the directional response at tens, hundreds, and thousands of \GeV. 
  \item \textbf{Numerical control and reproducibility.} We demonstrate convergence with steps-per-gyro and outer stop radius, provide figure-generation scripts and machine-readable tables, and define a default configuration that balances accuracy and runtime. The framework is lightweight and extensible, enabling future additions such as time dependence (solar-cycle evolution, CIRs/CMEs), PFSS-constrained polarity maps, and a geomagnetic-leg treatment for completeness.
\end{itemize}

\section{Magnetic-Field Models}
\label{sec:models}

We model the large-scale heliospheric magnetic field (HMF) with a set of analytic configurations of increasing realism and complexity. The goal is to capture the main geometric ingredients that control charge-sign drifts and azimuthal winding while keeping the forms compact and numerically stable for back-tracing. In each case we normalize to a prescribed radial field at 1~\AU, $\Br(1~\AU)$, and, unless stated otherwise, assume a steady solar wind. Complete formulae, sign conventions, and implementation details are collected in the Appendix. {\color{black}Note that the heliosphere extends beyond the nominal 50~AU boundary adopted in our back-tracing, encompassing the termination shock (TS) at $\sim$90~AU and the heliopause (HP) at $\sim$120~AU.  
We verified that extending the integration to 120~AU in representative Parker-field realizations changes the mean deflection by less than 0.05$^{\circ}$ at 100~GeV and less than 0.5$^{\circ}$ even at 10~GeV, confirming that the outer regions contribute negligibly to the overall bending.  
Although the magnetic-field intensity rises in the heliosheath, its scale length is large and its orientation remains nearly azimuthal, producing minimal additional curvature for incoming relativistic particles.  
A global north–south asymmetry of the heliosphere could in principle introduce small systematic offsets ($\lesssim$1$^{\circ}$), which we plan to incorporate in future work using numerical MHD field models.}

\subsection{Parker Spiral}
\label{subsec:parker}
The Parker model \citep{Parker1958} describes a magnetic field frozen into a steady, radial solar wind of speed $\Vsw$, wound into an Archimedean spiral by solar rotation $\Omega_\odot$. In heliocentric spherical coordinates $(r,\theta,\phi)$,
\begin{align}
\Br(r,\theta) &= A\,\Br(1~\AU)\,\Big(\frac{1~\AU}{r}\Big)^{\!2}, \\
\Bphi(r,\theta) &= -\,\frac{\Omega_\odot\, r \sin\theta}{\Vsw}\,\Br(r,\theta), \\
\Bth(r,\theta) &= 0,
\end{align}
where $A=\pm1$ encodes magnetic polarity. This baseline captures the dominant azimuthal winding and $r^{-2}$ decay and serves as our reference configuration (default parameters: $\Br(1~\AU)=\SI{5}{\nT}$, $\Vsw=\SI{400}{\kmps}$, $\Omega_\odot=\SI{2.865e-6}{rad\,s^{-1}}$). Flips across the heliospheric current sheet (HCS) may be treated by a sign change in $A$ across a specified neutral surface (see below).

\subsection{DQCS: Dipole + Quadrupole + Current Sheet}
\label{subsec:dqcs}
To imprint large-scale coronal topology (open polar fields and an equatorial streamer belt) we adopt the analytic dipole+quadrupole+current-sheet (DQCS) configuration of \citet{Banaszkiewicz1998}. In brief, a potential $\Phi(r,\theta)$ with dipolar and quadrupolar multipoles yields
\begin{equation}
\vb{B}_\text{PF} = -\grad \Phi(r,\theta),
\end{equation}
with a thin equatorial current-sheet term enforcing a polarity reversal between hemispheres. We parameterize the quadrupole-to-dipole ratio $Q$, a current-sheet fraction $\texttt{cs\_frac}$ (controlling the strength of the equatorial reversal), and a softening radius $r_0$ to regularize the field near the origin and around the sheet. The model is normalized so that the unsigned radial field at 1~\AU\ matches $\Br(1~\AU)$. Compared with Parker, DQCS introduces latitudinal structure in $\Br$ and a physically motivated neutral surface (flat HCS in the simplest limit).

\subsection{DQCS + Spiral (Smith--Bieber; Latitudinal Wind)}
\label{subsec:dqcs_spiral}
To reflect the observed azimuthal enhancement and latitudinal wind structure, we augment DQCS by adding a Parker-like spiral component and a Smith--Bieber–type strengthening of the azimuthal field \citep{SmithBieber1991}, together with a latitude-dependent wind speed $\Vsw(\theta)$ guided by \emph{Ulysses}-era measurements \citep{McComas2000}. Concretely, we retain $(\Br,\Bth)$ from DQCS and set
\begin{equation}
\Bphi(r,\theta) \;=\; -\,\frac{\Omega_\odot\, r \sin\theta}{\Vsw(\theta)}\,\Br(r,\theta)\,\Big[1 + k_{\rm SB}\,f_{\rm SB}(r,\theta)\Big],
\end{equation}
where $k_{\rm SB}$ is an order-unity Smith--Bieber factor and $f_{\rm SB}$ encodes a mild radial/latitudinal dependence (Appendix). For the wind we use a smooth two-speed profile,
\begin{equation}
\Vsw(\theta) \approx \Vsw^{\rm slow} + \frac{\Vsw^{\rm fast}-\Vsw^{\rm slow}}{2}\,\Big[1+\cos^{n}\!\theta\Big],
\end{equation}
with $\Vsw^{\rm slow}\!\sim\!350$--$500~\kmps$ at low latitudes and $\Vsw^{\rm fast}\!\sim\!650$--$800~\kmps$ toward the poles (default $n$ sets the transition width). This hybrid preserves the DQCS large-scale topology while introducing realistic azimuthal winding and latitudinal shear.

We also allow a \emph{wavy, tilted} HCS by specifying a neutral surface
\begin{equation}
\theta_{\rm HCS}(\phi) \;=\; \frac{\pi}{2} \;+\; \alpha \,\sin\!\big[m\,\phi + \phi_0\big],
\end{equation}
with tilt $\alpha$, wave number $m$ (typically $m=1$–$2$), and phase $\phi_0$, consistent with solar-cycle variability of the current sheet \citep{Smith2001HCS,WangSheeley1992}. Across this surface we reverse the sign of the radial field component, producing sector structure.

\subsection{Fisk-like Variant}
\label{subsec:fisk}
Finally, we incorporate a small latitudinal component to emulate Fisk-like fields, in which nonradial motion of footpoints at the Sun produces a finite $\Bth$ and modified spiral connectivity \citep{Fisk1996}. We model this as an $\mathcal{O}(\epsilon)$ perturbation to the Parker-like geometry,
\begin{equation}
\Bth(r,\theta) \;=\; \epsilon\,g(r,\theta)\,\Br(r,\theta),
\end{equation}
with $0\!\le\!\epsilon\!\ll\!1$ and a smooth $g$ chosen so that $\div\vb{B}\!\approx\!0$ to first order (see the Appendix). This prescription captures the principal directional effect—small systematic drifts associated with $\Bth\neq0$—without resorting to a full time-dependent treatment.

\medskip
Across all models we ensure continuity of $\vb{B}$ away from the HCS, enforce the correct $r^{-2}$ falloff of open flux, and re-scale the overall field strength via $\Br(1~\AU)$ to explore solar-cycle variability. Polarity $A$ and HCS geometry control charge-sign effects and sector crossings; these are scanned in the results that follow.

{\color{black}The magnetic-field models adopted in this work—the Parker spiral, the modified Parker field including latitudinal corrections, and the current-sheet–augmented configuration—are all consistent, within their nominal approximations, with in-situ measurements by spacecraft such as {\it Ulysses}, {\it Voyager}, and {\it Parker Solar Probe}.  
In particular, the field magnitude and polarity reversal across the heliospheric current sheet reproduce the observed trends with heliocentric distance and latitude out to $\sim50$ AU, beyond which the analytic forms become increasingly uncertain.  
While these models do not include temporal variations associated with the solar cycle or turbulence, they capture the large-scale geometry and intensity of the heliospheric field relevant for high-energy particle deflections.}

{\color{black}Throughout this work, we define the particle rigidity as $R \equiv p/|q| = \sqrt{E^{2}-m^{2}}/|q|$, where $p$ and $q$ are the particle momentum and charge, respectively.  
For relativistic electrons and positrons, $R \simeq E/|e|$, so that energy and rigidity are effectively interchangeable at the energies considered here.  
We retain the notation in terms of energy for clarity but emphasize that all magnetic deflections scale as $R^{-1}$.}

\section{Numerical Methods}
\label{sec:method}

We back-trace relativistic test particles from the observer at $r=1\,\AU$ through a prescribed, time-steady heliospheric field $\vb{B}(r,\theta,\phi)$ until an outer radius $r_{\rm stop}$ (typically 50--100~\AU) is reached, at which point the particle direction is recorded as the ``asymptotic'' direction. Throughout, we ignore electric fields in the plasma frame (ideal MHD) and treat particles as ultrarelativistic with fixed speed $v\simeq c$. The equations of motion are
\begin{align}
\frac{d\vb{x}}{dt} &= \vb{v},\\
\frac{d\vb{p}}{dt} &= q\,\vb{v}\times \vb{B}(\vb{x}),
\end{align}
with $\vb{p}=\gamma m\vb{v}$ and $\gamma$ the Lorentz factor. In a purely magnetostatic field, $|\vb{p}|$ is conserved; any small numerical drift is removed by renormalizing $|\vb{v}|$ to $c$ every substep.

\paragraph{Back-tracing convention.} We integrate \emph{backwards} by evolving the equations above with the measured (local) initial direction at $1\,\AU$ but with the sign of the charge flipped ($q\to -q$). This is equivalent to reversing time in a magnetic field and is numerically convenient.

\paragraph{Integrator and timestep.} We use a classical fourth-order Runge--Kutta (RK4) scheme (fixed stage coefficients) with a step size tied to the local relativistic gyroperiod,
\begin{equation}
\Omega_{\rm g} \equiv \frac{|q|\,|\vb{B}|}{\gamma m c},\qquad
\Delta t \equiv \frac{2\pi}{\Omega_{\rm g}}\;\frac{1}{N_{\rm gyr}},
\end{equation}
where $N_{\rm gyr}\equiv\texttt{steps\_per\_gyr}$ is user-controlled (default~1,500). Thus $\Delta t \propto 1/|\vb{B}|$ adapts mildly in space. To avoid excessively small steps near sharp current-sheet transitions, we (i) smooth the HCS sign change over a narrow angular width $\delta$ (Appendix~\ref{app:hcs}) and (ii) impose practical caps on $\Delta t$ and the per-step spatial advance. The RK4 choice is robust and sufficiently accurate for the smooth analytic fields used here \citep[e.g.,][]{Press2007,DormandPrince1980}.

\paragraph{Initial conditions.} The observer is placed at $(r,\theta,\phi)=(1\,\AU,\theta_{\rm obs},\phi_{\rm obs})$; unless otherwise stated we adopt $\theta_{\rm obs}\!\approx\!\pi/2$ (ecliptic) and a reference longitude $\phi_{\rm obs}=0$. A local direction $(\theta_{\rm loc},\phi_{\rm loc})$ relative to the spherical basis $(\hat{\vb r},\hat{\boldsymbol\theta},\hat{\boldsymbol\phi})$ is converted to a Cartesian unit vector $\hat{\vb u}_0$ and scaled to $v\simeq c$.

\paragraph{Asymptotic direction and deflection.} We stop when $r\ge r_{\rm stop}$ (default 50~\AU; checked up to 100~\AU) and record the unit velocity $\hat{\vb u}_{\infty}$. The heliospheric deflection angle for that trajectory is
\begin{equation}
\Delta\theta \;=\; \arccos\!\big(\hat{\vb u}_0\!\cdot\!\hat{\vb u}_{\infty}\big)\,,
\end{equation}
which we then aggregate across ensembles. 

\paragraph{Directional sampling.} For a given kinetic energy $E$, we sample $N$ arrival directions isotropically on the unit sphere by drawing $u\sim \mathcal{U}(0,1)$ and $v\sim \mathcal{U}(0,1)$ and setting
\[
\cos\theta_{\rm loc}=2u-1,\qquad \phi_{\rm loc}=2\pi v,
\]
with a fixed seed for reproducibility. We report the mean, median, standard deviation, and 16--84\% / 5--95\% percentiles of $\Delta\theta$ as a function of $E$.

\paragraph{Convergence controls.} We verify numerical stability by scanning over $N_{\rm gyr}$ and $r_{\rm stop}$ (Fig.~\ref{fig:numstab_steps} and \ref{fig:stab_bound}, respectively). In all configurations shown in the main text, statistics of $\Delta\theta$ are stable to better than the line thickness once $N_{\rm gyr}\!\gtrsim\!1,500$ and $r_{\rm stop}\!\gtrsim\!50\,\AU$.

\paragraph{Energy and units.} The Lorentz factor is $\gamma=1+E/(mc^2)$, with $m$ the electron mass and $q=\pm e$. We adopt SI internally and provide inputs as $\Br(1\,\AU)$ in \nT, $\Vsw$ in \kmps, and $E$ in \GeV. The field models are normalized such that $\langle|\Br(1\,\AU)|\rangle_{\Omega}$ equals the requested value (Appendix~\ref{app:norm}).

{\color{black}As a numerical consistency check, we verified that for particle rigidities above 50 TV (corresponding to energies $\gtrsim50$ TeV), the computed mean angular deflection falls below $10^{-3}$ rad, confirming that the integration accuracy is sufficient to reproduce the expected ballistic limit.  
This test defines the numerical “floor” below which computed deflections are dominated by finite-precision effects rather than physical bending.  
The code was also validated against analytic test cases involving uniform magnetic fields, where trajectories are recovered within $<0.1^{\circ}$ of the exact helical solution.}

\subsection{Numerical Stability and Asymptotics}
\label{sec:numerics}

\begin{figure*}[t]
  \centering
  \begin{subfigure}[t]{0.49\textwidth}
    \centering
    \includegraphics[width=\linewidth]{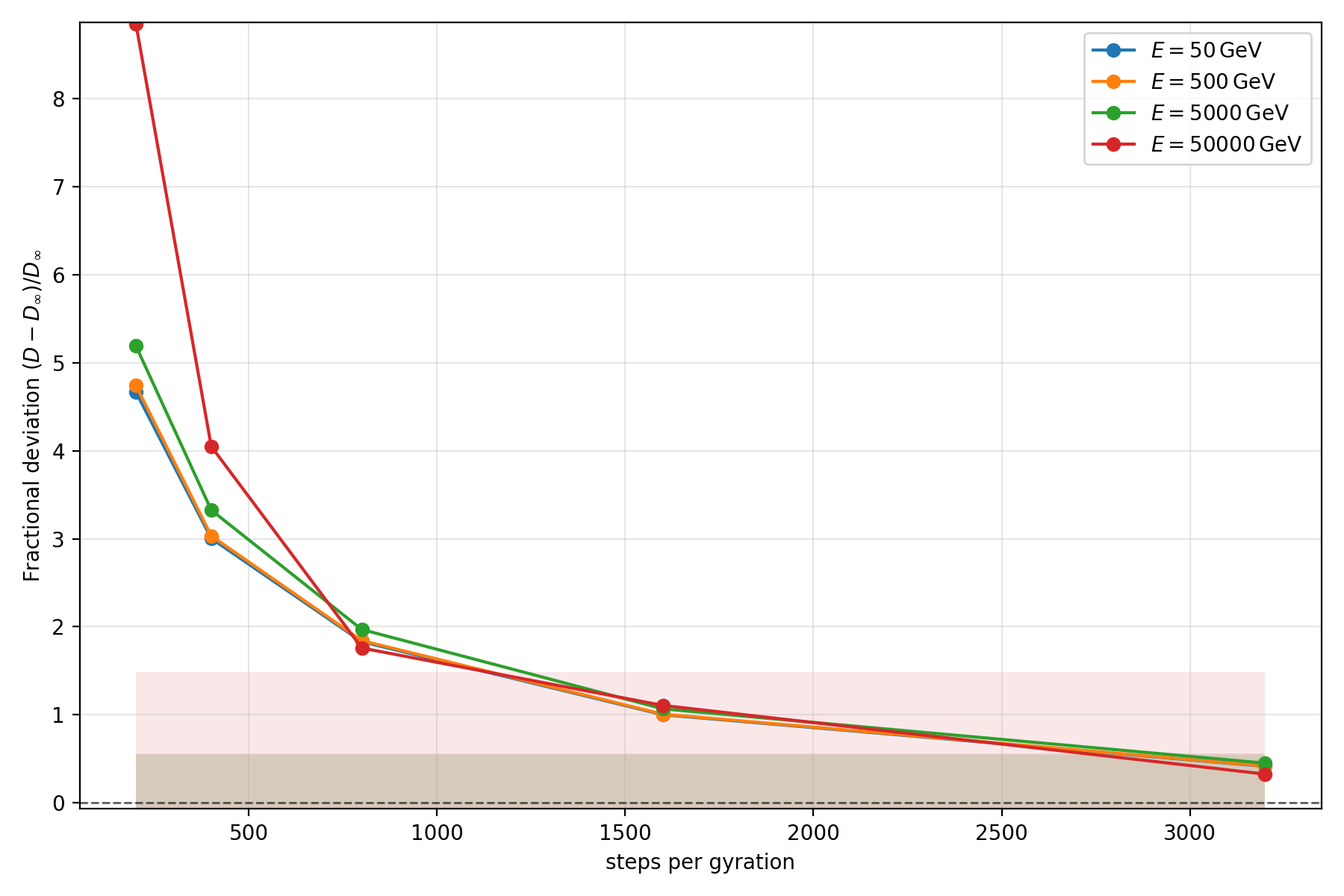}
    \caption{Fractional numerical error}\label{fig:numstab_steps}
  \end{subfigure}\hfill
  \begin{subfigure}[t]{0.49\textwidth}
    \centering
    \includegraphics[width=\linewidth]{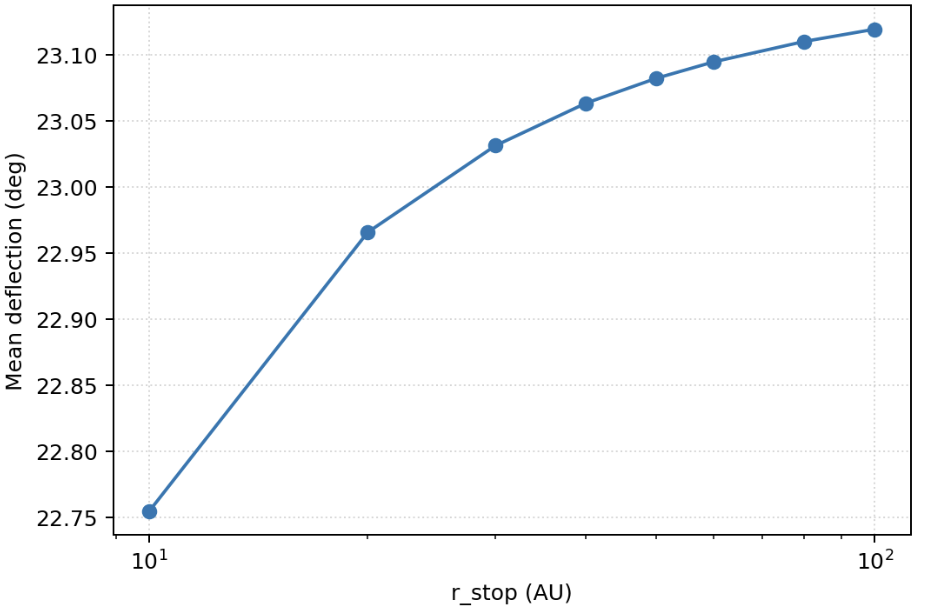}
    \caption{Mean deflection versus outer boundary $r_{\rm stop}$}
    \label{fig:stab_bound}
  \end{subfigure}
\caption{(a): Fractional numerical error on the sky--averaged deflection for the Parker unipolar reference model at fixed outer radius $r_{\rm stop}=50$\,AU.
  We show $(D - D_\infty)/D_\infty$ as a function of the number of integration steps per gyration for four representative energies ($E = 50, 500, 5\times10^3$, and $5\times10^4$~GeV), where $D_\infty$ is estimated from the highest–resolution run (${\rm steps/gyro}=3200$).
  All curves are computed with 64 deterministic sky directions.
  The shaded horizontal bands mark target tolerances of $|\Delta D|/D_\infty <0.5\%$ (dark band) and $<1.5\%$ (light band).
  For ${\rm steps/gyro}\gtrsim 1300$ the fractional deviation for all energies lies well within the $1.5\%$ band, demonstrating stable convergence of the integrator at the resolutions adopted for our production runs. (b): mean deflection versus outer boundary $r_{\rm stop}$ at fixed resolution.
  Configuration: DQCS$+$spiral with a wavy HCS ($m=1$, tilted–sheet polarity), $N=100$ random sky directions per point.
  Error bars are omitted for clarity; trends are robust to the number of rays.}
\end{figure*}

We discuss here numerical convergence tests we carried out on our model implementation. In particular, the numerical tests in Fig.~\ref{fig:numstab_steps}  quantify the robustness of our trajectory integration scheme and complement the model--comparison figures discussed above.  In the left panel, we focus on the Parker unipolar field and track, for several fixed energies, and show how the sky--averaged deflection approaches an asymptotic value as the integration step budget is increased.  Expressing the result as the fractional offset $(D-D_\infty)/D_\infty$ relative to the highest--resolution run makes the convergence behavior explicit: for each energy the curves fall monotonically toward zero, and for ${\rm steps/gyro}\gtrsim1600$ the residual error is below the $1.5\%$ band (and typically at the $\sim 1\%$ level or better).  This demonstrates that our production choice (which is more conservative than the minimum satisfying these criteria) yields deflections that are numerically converged for the purposes of our comparison across magnetic geometries.

Fig.~\ref{fig:stab_bound} addresses a distinct question: the sensitivity of the mean deflection to the location of the outer integration boundary, for a representative structured field (DQCS+spiral with a wavy HCS).  Here we hold the time step fixed and vary $r_{\rm stop}$, finding that the mean deflection increases only marginally once the integration extends beyond $\sim 40$--50~AU and effectively saturates by $r_{\rm stop}\simeq 50$~AU.  The residual change between 50 and 100~AU is at the sub-percent level, confirming that the dominant contribution to heliospheric bending arises in the inner heliosphere where the large-scale field is strongest and the Parker winding is most pronounced.  Together, these tests validate our baseline numerical settings and ensure that the trends reported in the previous sections are controlled by physical input (field geometry and charge sign) rather than by integration artifacts.

\section{Results}
\label{sec:results}

\begin{figure}[t]
  \centering
  \begin{subfigure}[t]{0.49\linewidth}
    \centering
    \includegraphics[width=\linewidth]{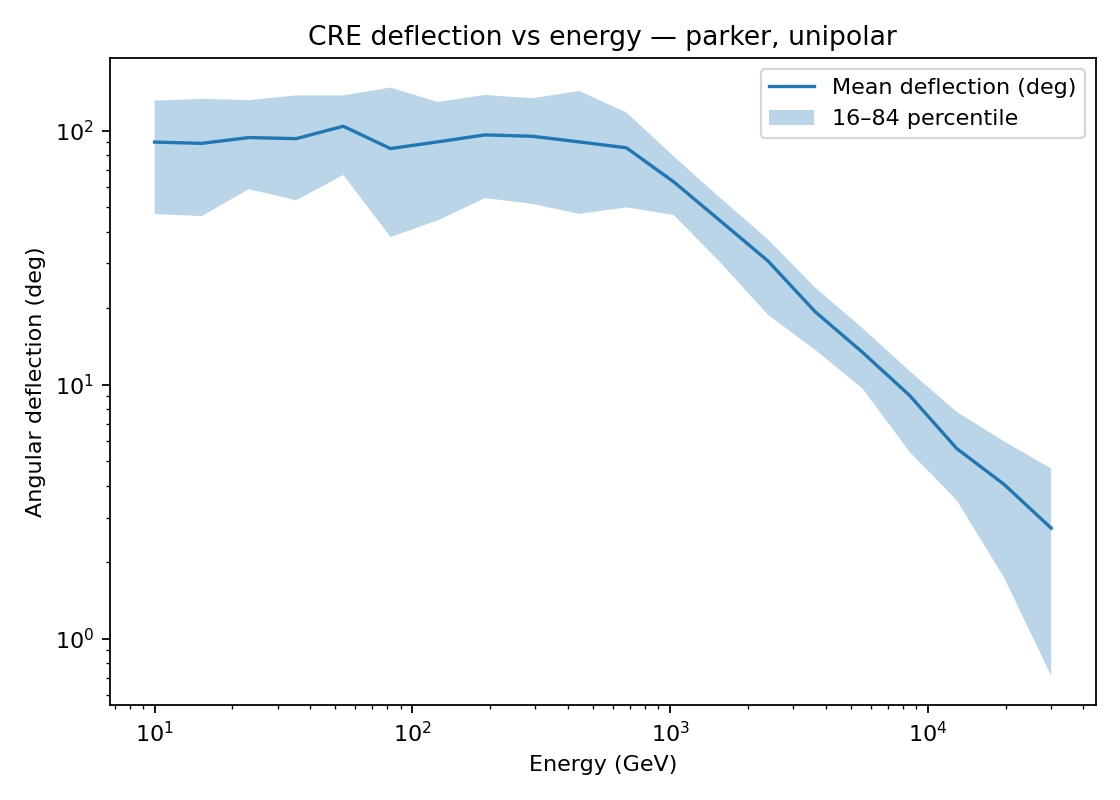}
    \caption{Electrons ($q=-e$).}
    \label{fig:parker_electron}
  \end{subfigure}
  \hfill
  \begin{subfigure}[t]{0.49\linewidth}
    \centering
    \includegraphics[width=\linewidth]{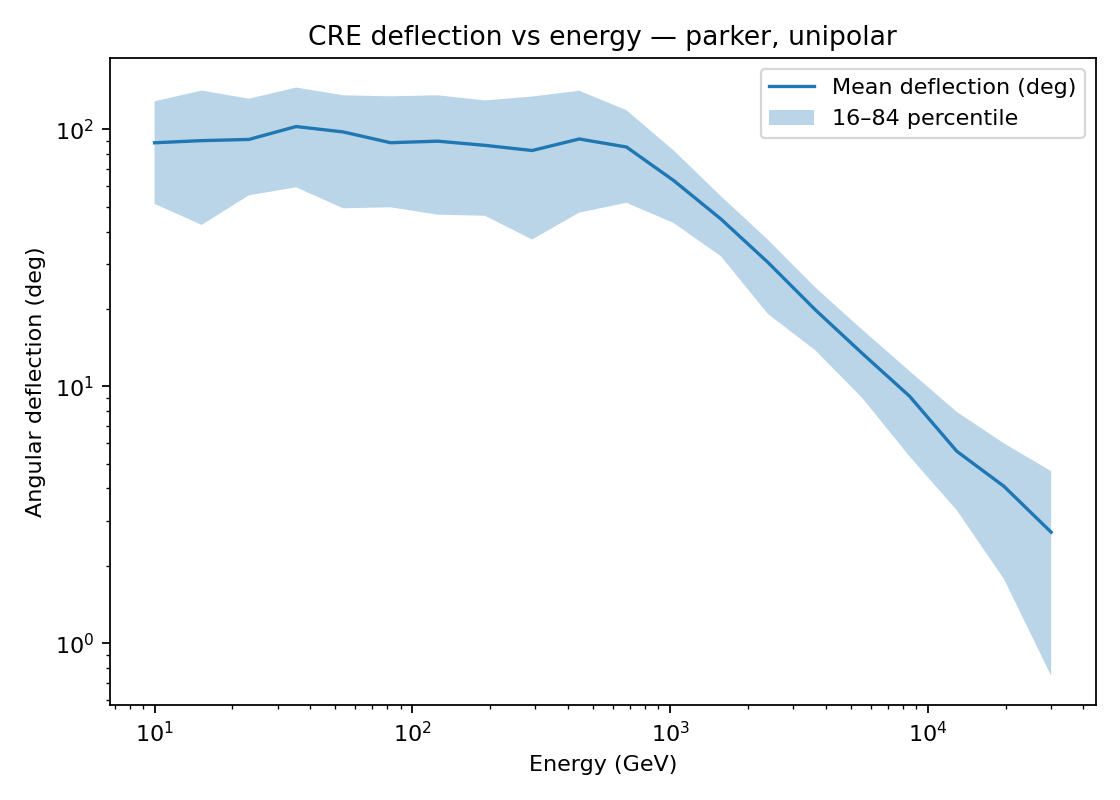}
    \caption{Positrons ($q=+e$).}
    \label{fig:parker_positron}
  \end{subfigure}
  \caption{Baseline Parker spiral with $\Br(1\,\AU)=5$\,nT, $\Vsw=400$\,km\,s$^{-1}$, 
  $\Omega_\odot=2.865\times10^{-6}$\,rad\,s$^{-1}$. 
  Each panel shows the mean heliospheric deflection $\langle\Delta\theta\rangle$ (solid line) 
  and the 16--84\% percentile band (shaded) over 100 isotropically sampled arrival directions, 
  as a function of energy from 10~GeV to 30~TeV. {\color{black} Note that at energies $\lesssim$10~GeV), the effective dispersion is limited by our back-tracing cutoff at 50~AU, as discussed in the text.}}
  \label{fig:deflection_vs_energy_parker}
\end{figure}

\subsection{Baseline: Deflection vs Energy (Parker)}
\label{sec:results_parker}

Under the canonical Parker-spiral configuration \citep{Parker1958} with a steady, spherically symmetric solar wind, the mean angular deflection decreases rapidly with energy. Figures~\ref{fig:parker_electron} and \ref{fig:parker_positron} show electrons and positrons separately; both exhibit an approximate $\Delta\theta\!\propto\!E^{-1}$ trend once $E\gtrsim\,$few$\times10^2$~GeV, as expected from the gyrofrequency scaling $\Omega_{\rm g}\propto |B|/\gamma$. 

At tens of GeV, typical deflections are $\mathcal{O}(10^2~\mathrm{deg})$, implying that heliospheric bending largely scrambles arrival directions. To make this statement concrete: if heliospheric bending is strong enough that the outgoing direction is effectively randomized on the sky, the expected angular offset between the true and observed directions is
\(\langle \Delta\theta \rangle = 90^\circ\).
This follows because the separation \(\Delta\theta\) between two independent isotropic directions on the sphere has probability density
\(p(\Delta\theta)=\tfrac{1}{2}\sin\Delta\theta\) for \(\Delta\theta\in[0,\pi]\), whose mean and median are both \(\pi/2=90^\circ\).
Thus, when our predicted deflections approach tens of degrees and trend toward isotropy, ``scrambled'' arrival directions should be interpreted as \emph{order–unity} loss of directional information, with a characteristic offset of \(\sim 90^\circ\). {\color{black}At low energies ($\lesssim$10~GeV), the effective dispersion is limited by our back-tracing cutoff at 50~AU: many trajectories become chaotic within the inner heliosphere but are terminated before achieving full isotropy, resulting in an apparent confinement of the angular spread to $\sim\!\pm50^{\circ}$.  
This does not imply physical focusing, but rather reflects the finite integration volume adopted here.  
Conversely, at the highest energies ($\gtrsim$ a few~TeV), the apparent growth of the dispersion is a numerical artifact caused by the loss of significant digits in the determination of nearly parallel trajectories; physically, the distribution should converge to a narrow peak with $\sigma_{\Delta\theta}\!\to\!0$.}

{\color{black}Numerically, we found that the mean angular deflection $\langle\Delta\theta\rangle$ exhibits a plateau at energies greater than 50 TeV, hence to the right of the range shown in the figure, and depending on the adopted HMF configuration.  This plateau arises because, at higher energies,  numerical noise approaches the integration precision limit; results in this regime should therefore not be interpreted physically, and are not shown in our figures.}

The 16--84\% band narrows with energy, reflecting the decreasing sensitivity to sky location as the Larmor radius grows relative to heliospheric length scales. 

Charge-sign differences are modest in this unipolar baseline (no tilted/wavy HCS), but they are nonzero at sub-TeV energies due to drifts in the azimuthal component $B_\phi$. These baseline curves set reference thresholds $E_{\rm crit}(\theta_{\rm inst})$ for later comparisons: given an instrument’s angular budget $\theta_{\rm inst}$, one can read off the energy above which heliospheric deflection is negligible relative to the point-spread function. Subsequent sections explore how these thresholds shift when we introduce a wavy/tilted current sheet, a latitudinal wind and Smith--Bieber enhancement, Fisk-like $B_\theta$ perturbations, and variations of $\Br(1\,\AU)$.



\begin{figure*}[t]
  \centering
  \begin{subfigure}[t]{0.49\textwidth}
    \centering
    \includegraphics[width=\linewidth]{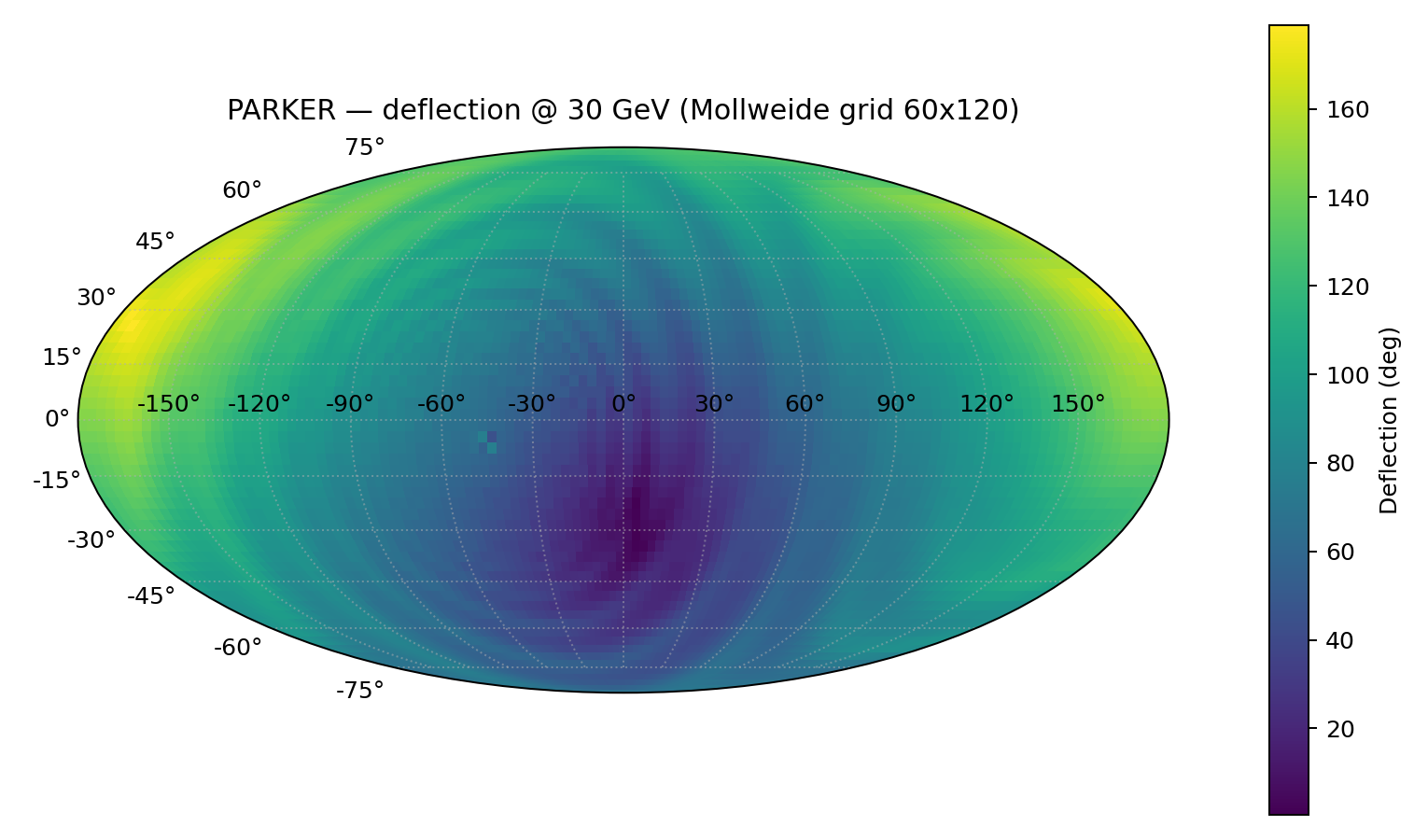}
    \caption{$E=30$ GeV}
  \end{subfigure}
  \hfill
  \begin{subfigure}[t]{0.49\textwidth}
    \centering
    \includegraphics[width=\linewidth]{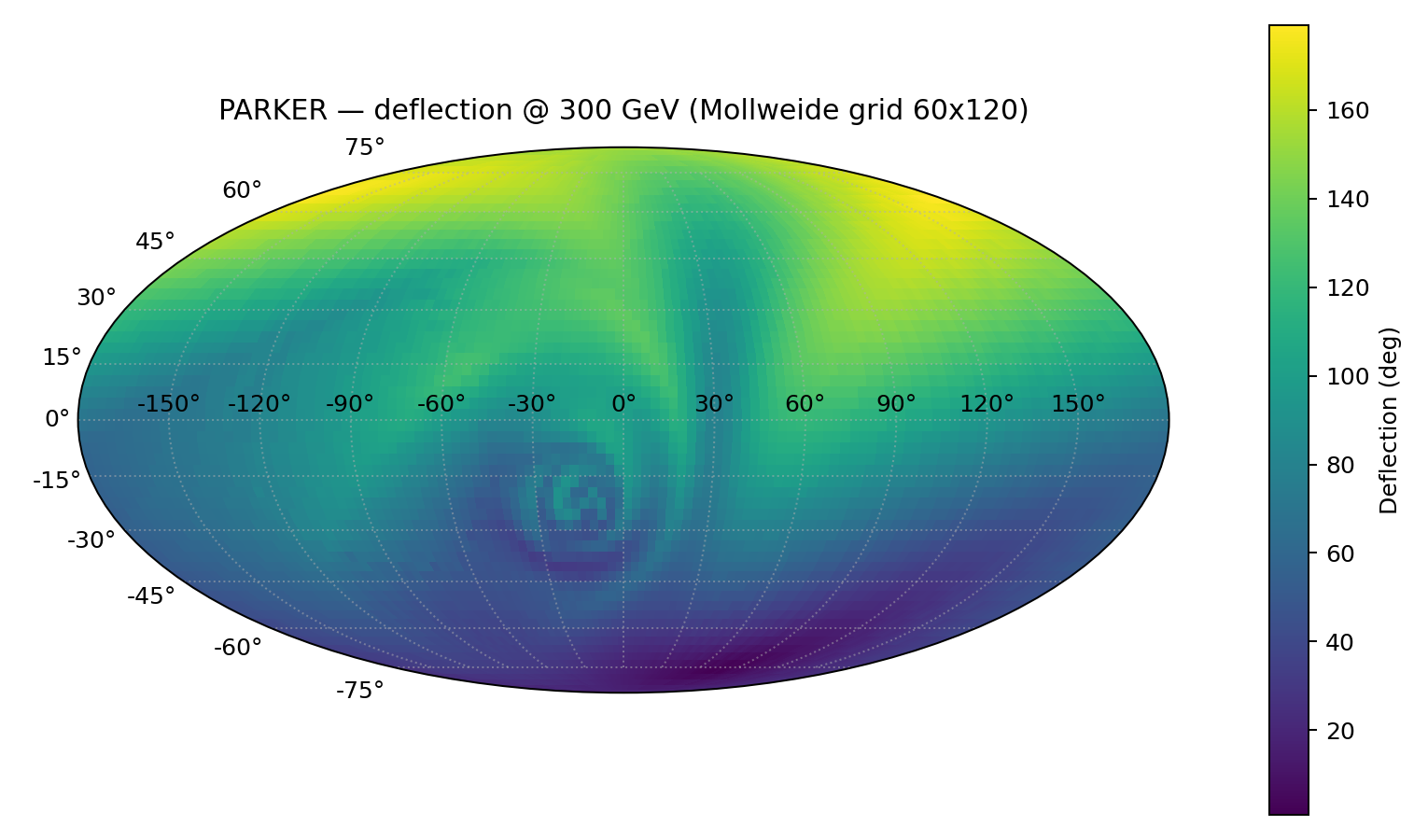}
    \caption{$E=300$ GeV}
  \end{subfigure}\\[0.6em]
  \begin{subfigure}[t]{0.49\textwidth}
    \centering
    \includegraphics[width=\linewidth]{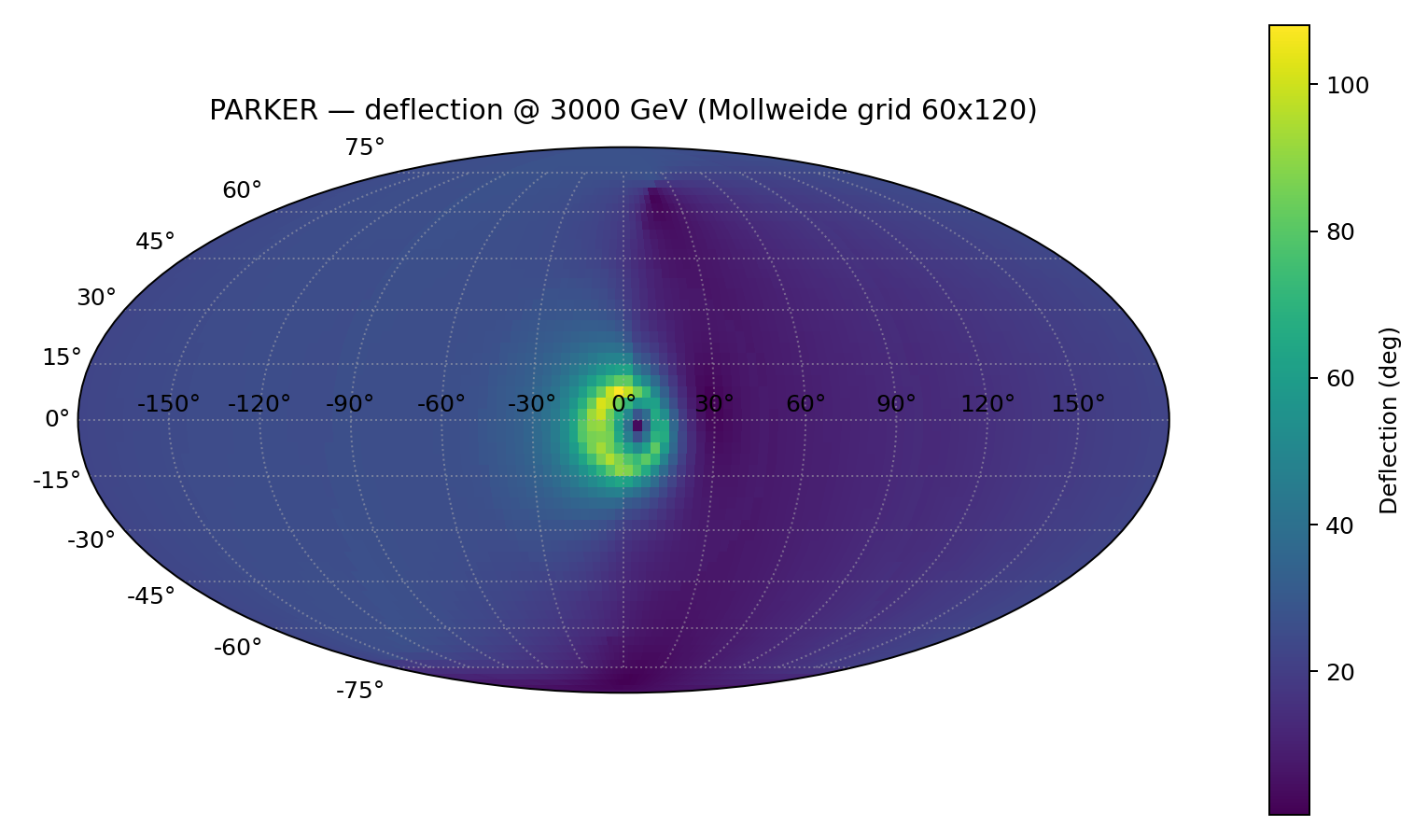}
    \caption{$E=3$ TeV}
  \end{subfigure}
  \hfill
  \begin{subfigure}[t]{0.49\textwidth}
    \centering
    \includegraphics[width=\linewidth]{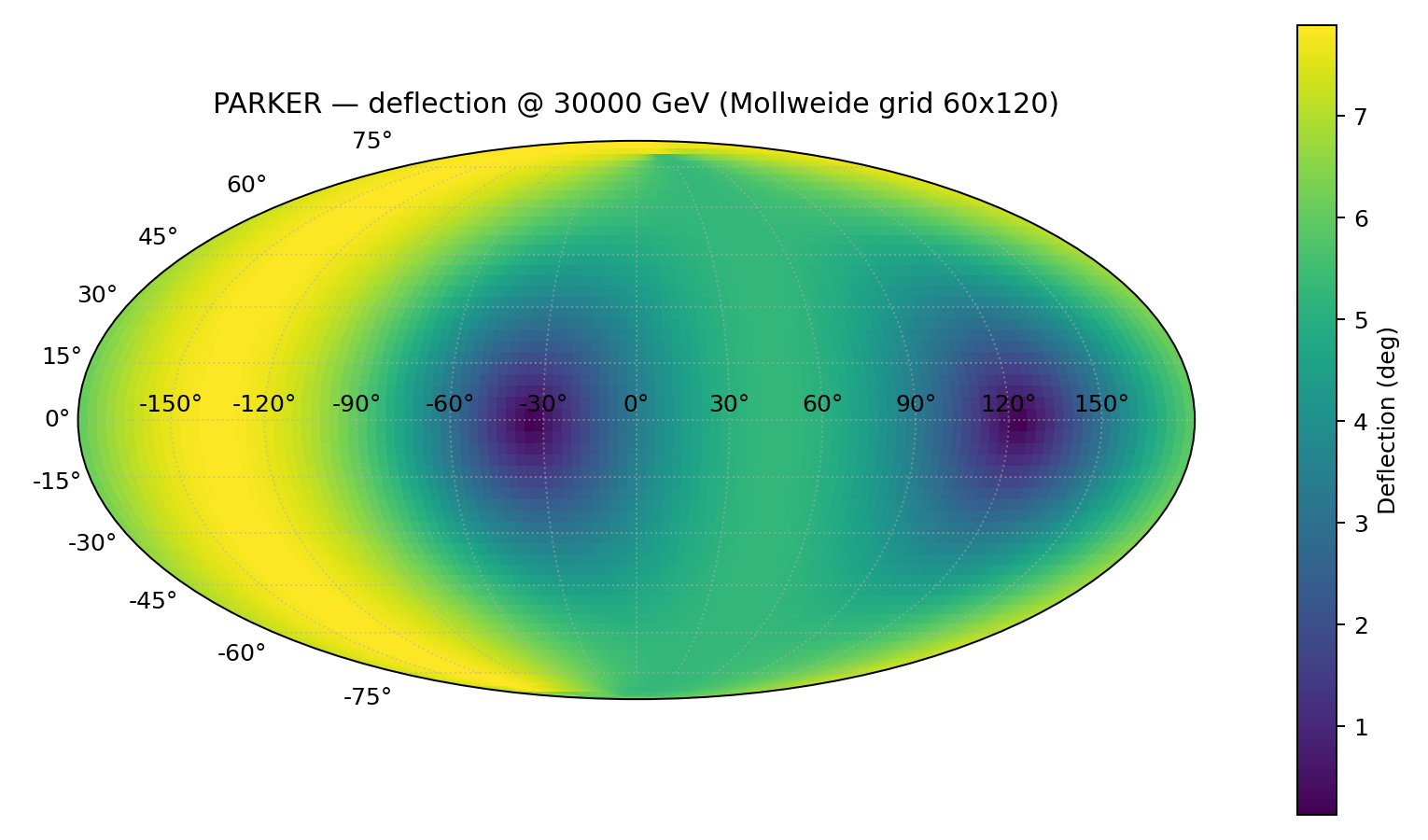}
    \caption{$E=30$ TeV}
  \end{subfigure}
  \caption{Sky maps of the heliospheric deflection angle (degrees; Mollweide projection, $60\times120$ grid) for the baseline Parker spiral with $\Br(1\,\AU)=5$ nT, $\Vsw=400$ km s$^{-1}$, and unipolar polarity. Each panel shows the angular deviation between the local arrival direction at $1$ AU and the asymptotic direction at $r_{\rm stop}=50$ AU for electrons, computed direction–by–direction with identical numerical settings (steps-per-gyro $=80$). Color scales are per–panel and labeled in degrees -- note especially the significantly different scale for the bottom, right panel. {\color{black} Note that the maps are displayed in heliocentric ecliptic coordinates, with the $(0,0)$ direction corresponding to the observer’s line of sight (i.e., the Sun–Earth axis), see the text for details.}}
  \label{fig:skymaps_parker}
\end{figure*}

\subsection{Sky Anisotropy of Bending}
\label{sec:sky_anisotropy}

Figure~\ref{fig:skymaps_parker} visualizes the spatial structure of heliospheric bending in the baseline Parker spiral across four representative energies. The maps display the angular deflection $\Delta\theta(\hat{\boldsymbol{n}})$ for each line of sight $\hat{\boldsymbol{n}}$ on the sky, obtained by back–tracing from $1$ AU to $r_{\rm stop}=50$ AU with identical numerical controls as in the energy–sweep results. {\color{black}The maps in Fig.~\ref{fig:skymaps_parker} are displayed in heliocentric ecliptic coordinates, with the $(0,0)$ direction corresponding to the observer’s line of sight (i.e., the Sun–Earth axis).  
Deflections are measured relative to this axis, such that particles arriving undeflected appear near the map center.  
Positive longitudes correspond to solar rotation direction, while latitudes increase toward the north heliographic pole.  
In the high-energy limit ($E\!\gtrsim\!1$~TeV), the maps indeed converge to a compact region centered around $(0,0)$, confirming that deflections vanish as expected.}

Four systematic trends emerge:

\emph{(i) Global amplitude and energy scaling.} The overall magnitude of the deflection decreases rapidly with energy, consistent with the expectation that the effective curvature imparted by the large–scale field scales as $\propto |B|/E$. At $E=30$ GeV the deflection is broadly of order $10^2$ degrees over much of the sky, implying that arrival directions are effectively scrambled by the heliospheric leg. By $E=300$ GeV a large–scale gradient remains, but sub–degree patches begin to emerge. At $E=3$--$30$ TeV the map contracts into a compact region of enhanced bending embedded within a mostly low–deflection sky, mirroring the shrinking influence of the azimuthal spiral as the Larmor radius grows.

\emph{(ii) Longitudinal asymmetry aligned with the Parker spiral.} The anisotropy is not purely latitudinal. Instead, a clear longitudinal structure aligns with the local tangent of the Parker spiral at the observer: a crescent– or annulus–like feature appears near the longitude where the line of sight is close to the local field direction, surrounded by a ring where small changes in launch direction sample significantly different path curvatures. This morphology is most visible at $E=300$ GeV and sharpens into a compact “lens” at $E=3$ TeV. Physically, trajectories launched nearly parallel to the local field experience minimal instantaneous $v\times B$ bending near $1$ AU and retain similar pitch angles as they exit the inner heliosphere, whereas those launched slightly offset accumulate larger azimuthal rotation and cross to neighboring spiral surfaces, producing the ring of enhanced $\Delta\theta$.

\emph{(iii) Equatorial versus polar behavior.} A broad latitudinal gradient persists at all energies: directions closer to the ecliptic/heliographic equator differ systematically from higher latitudes. In the Parker geometry, the azimuthal component scales as $B_\phi\propto \sin\theta$ and the radial field as $B_r\propto r^{-2}$; combined with the different chord lengths that rays traverse through regions of strong $|B_\phi|$, this produces a measurable equatorial–polar contrast. At low energies the contrast is gentle and spread over large angular scales; at multi–TeV energies, the contrast is expressed mainly through the compact lens and surrounding ring, with the remainder of the sky nearly uniformly at small deflection.

\emph{(iv) ``Hot spots''.} The bright ``hot spots’’ are not true sources but geometric caustics of the Parker–spiral mapping. They occur where the line of sight is nearly tangent to the local spiral at 1~AU, so neighboring launch directions sample markedly different path curvatures: rays aimed slightly off the tangent accrue larger $v\times B$ rotation and cross adjacent spiral surfaces, producing a ring or crescent of enhanced $\Delta\theta$. This feature sits near the ecliptic because $B_\phi\propto \sin\theta$ maximizes there, and it tightens with energy: at $E=30$~GeV the caustic smears into a broad band, by $E=300$~GeV it becomes a distinct annulus, and by $E=3$--$30$~TeV it collapses into a compact lens embedded in a largely low–deflection sky. The longitude of the hot spot tracks the local field direction at the observer and would shift under HCS tilt/waviness or DQCS–like topology; its handedness flips between electrons and positrons. Accordingly, down–weighting these regions can stabilize anisotropy estimates at sub–TeV energies without biasing high–energy conclusions. {\color{black}The progressive shift of the focusing region with increasing energy arises because the average path length through the heliospheric magnetic field decreases as particle rigidity grows, causing particles to sample magnetic-field lines at progressively smaller heliocentric latitudes and longitudes.  
For a Parker-like field, the tangent direction of the spiral at 1~AU deviates from the radial direction by an angle $\tan^{-1}(\Omega r_{\odot}/v_{\rm sw})\simeq45^{\circ}$ for a solar-wind speed of $v_{\rm sw}=400$~km~s$^{-1}$.  
As $r_{\rm L}$ increases with energy, the effective deflection vector aligns more closely with this local field orientation, shifting the centroid of the focusing region by $\Delta\phi\simeq10^{\circ}$–$20^{\circ}$ between 10~GeV and 1~TeV, consistent with the trends seen in Fig.~\ref{fig:skymaps_parker}.}

{\color{black}The “lens” or focusing feature predicted in our maps has not yet been clearly identified in existing cosmic-ray electron or positron anisotropy observations.  
Current measurements by HAWC, AMS-02, and DAMPE lack either the required angular resolution or statistics at sub-TeV energies to isolate such small-scale heliospheric signatures.  
A full treatment of transport, including diffusion and drift, would be expected to smooth and partially wash out this feature, especially below $\sim50$~GeV, where particle trajectories become strongly stochastic.  
Nevertheless, at higher energies where the propagation is quasi-ballistic, the geometric imprint of the heliospheric field may remain detectable as a mild dipole-like distortion, motivating future high-statistics anisotropy searches.}

These maps provide a directional complement to the energy–only summary in Fig.~\ref{fig:deflection_vs_energy_parker}. In practice, they have two immediate implications for anisotropy studies. First, they delimit sky regions where heliospheric bending could most strongly smear or re–orient a Galactic anisotropy signal at a given energy; masking or down–weighting those regions can make upper limits more robust. Second, the compact, energy–dependent lens near the local spiral tangent implies a potential, mild rotation of any large–scale dipole reconstructed on the sky, with a sign that flips for positrons relative to electrons in sectors where the polarity changes (we quantify charge–sign and sector effects in Sec.~\ref{sec:model_comparisons}). 

{\color{black}For completeness, we ran similar skymaps for the different, more realistic models discussed in the following sections. While a few small differences arise, we found that no substantive or qualitative differences arise, and we therefore do not include the corresponding plots (which are available from the authors upon request). In particular, we found that relative to the pure Parker case, other models yielded similar large-scale patterns but exhibited enhanced north–south asymmetries near the current-sheet latitude, especially below $\sim100$~GeV.  
The inclusion of the current sheet also introduces small displacements of the focusing region across hemispheres, consistent with the expected polarity reversal of the field.  
At multi-TeV energies, all models converge to the same near-undeflected morphology, confirming that geometric differences are suppressed in the ballistic limit.}

{\color{black}An alternative representation of the same information could be obtained by constructing pitch-angle maps centered on the local heliospheric magnetic-field direction at 1~AU.  
Such maps would explicitly show how deflections distribute around the field line tangent and would facilitate comparison with magnetospheric back-tracing analyses.  
In our present analysis, this correspondence can be inferred from the displacement of the focusing region relative to the $(0,0)$ direction in Fig.~3; we defer a full pitch-angle mapping to a dedicated follow-up study.}

Finally, the clean, coherent morphology here is a direct consequence of the unipolar Parker baseline. Introducing a tilted/wavy heliospheric current sheet, a latitudinal wind and Smith–Bieber enhancement, or small $B_\theta$ perturbations (Fisk–like) will (i) shift the location of the lens feature relative to the ecliptic/equatorial frame, (ii) imprint additional azimuthal structure and broaden the ring, and (iii) enhance charge–sign differences—effects we explore in the following sections.


\begin{figure*}[!h]
  \centering
  \begin{subfigure}[t]{0.49\textwidth}
    \centering
    \includegraphics[width=\linewidth]{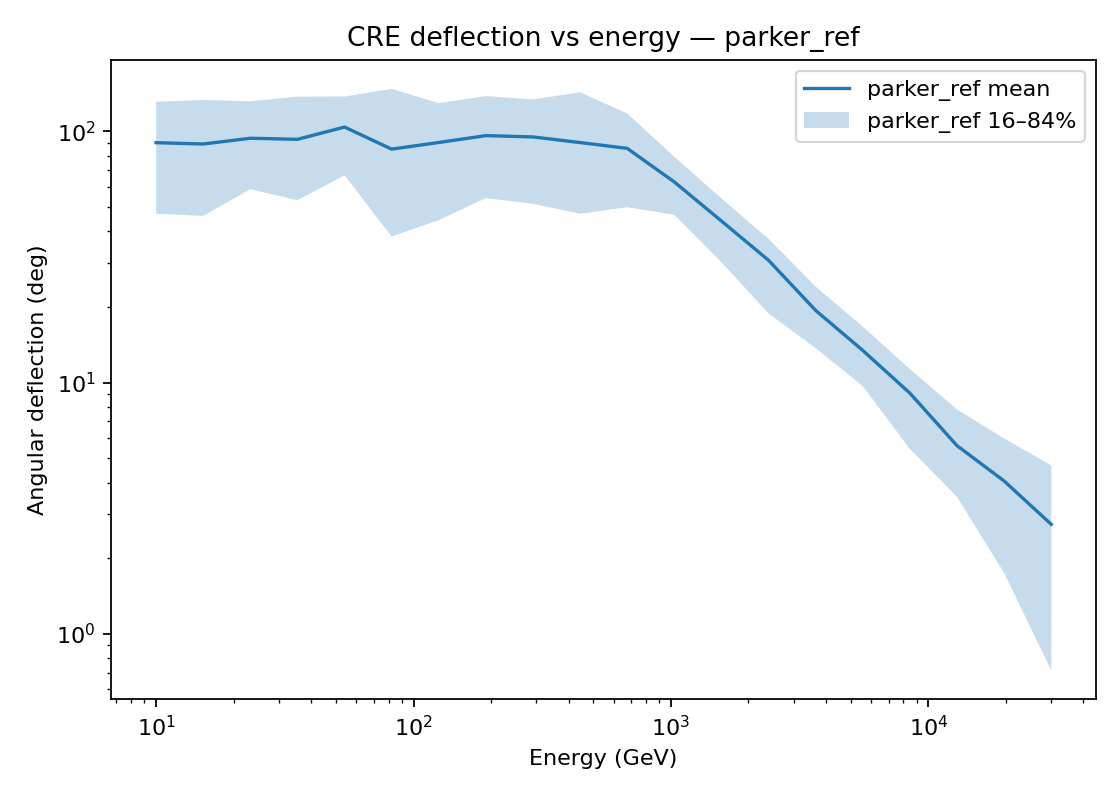}
    \caption{Parker baseline.}
    \label{fig:models_parker}
  \end{subfigure}\hfill
  \begin{subfigure}[t]{0.49\textwidth}
    \centering
    \includegraphics[width=\linewidth]{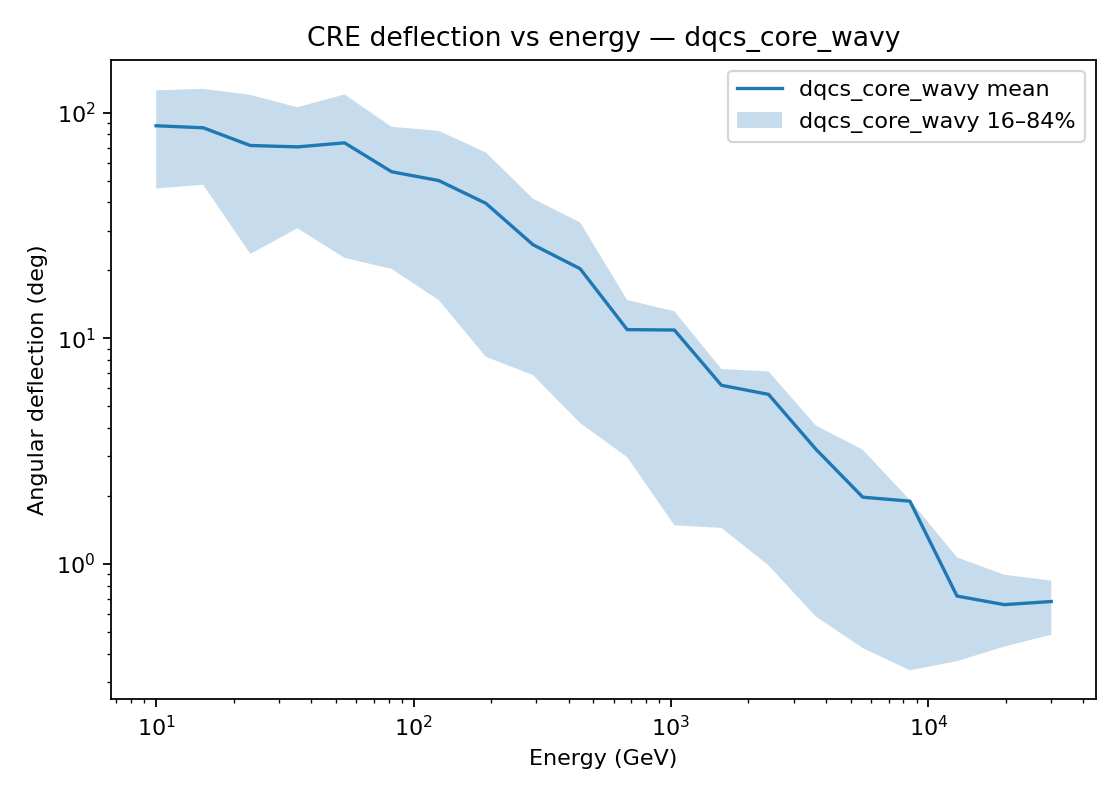}
    \caption{DQCS core only, wavy HCS sign (no $B_\phi$).}
    \label{fig:dqcs_core_wavy}
  \end{subfigure}\\[0.6em]
  \begin{subfigure}[t]{0.49\textwidth}
    \centering
    \includegraphics[width=\linewidth]{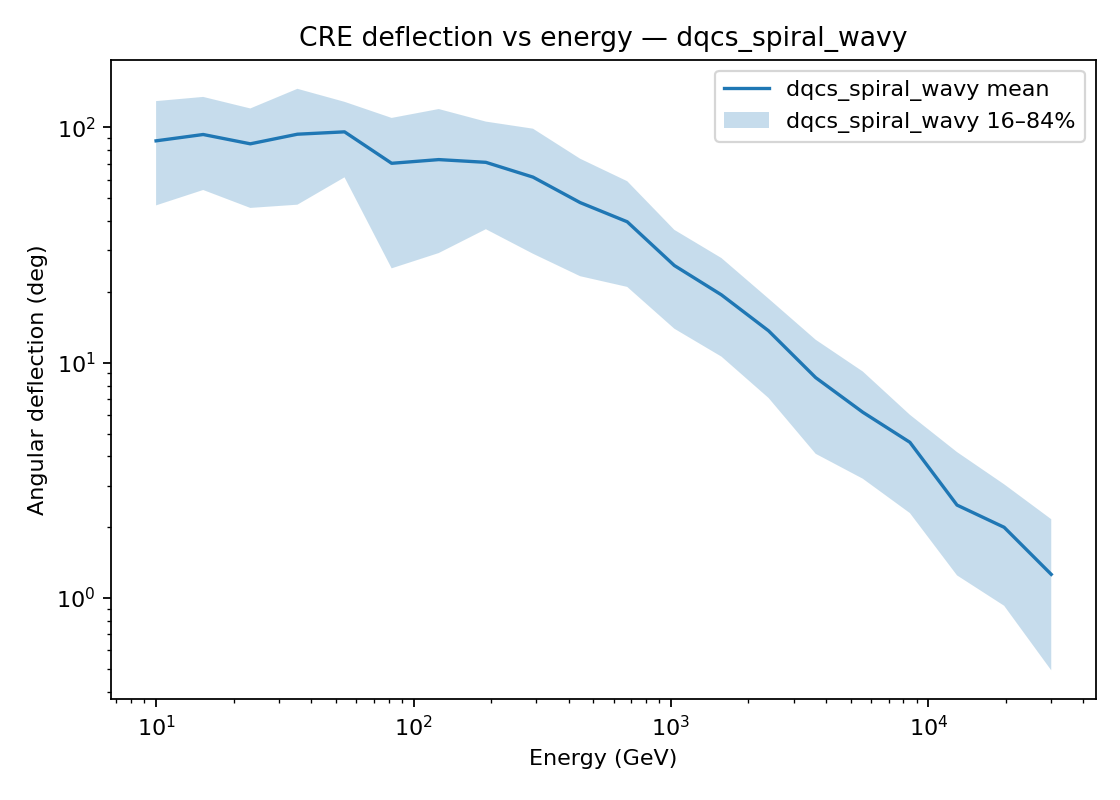}
    \caption{DQCS\,+\,spiral ($B_\phi$) with Smith--Bieber and latitudinal solar wind speed $V_{\rm sw}$; wavy HCS.}
    \label{fig:dqcs_spiral_wavy}
  \end{subfigure}\hfill
  \begin{subfigure}[t]{0.49\textwidth}
    \centering
    \includegraphics[width=\linewidth]{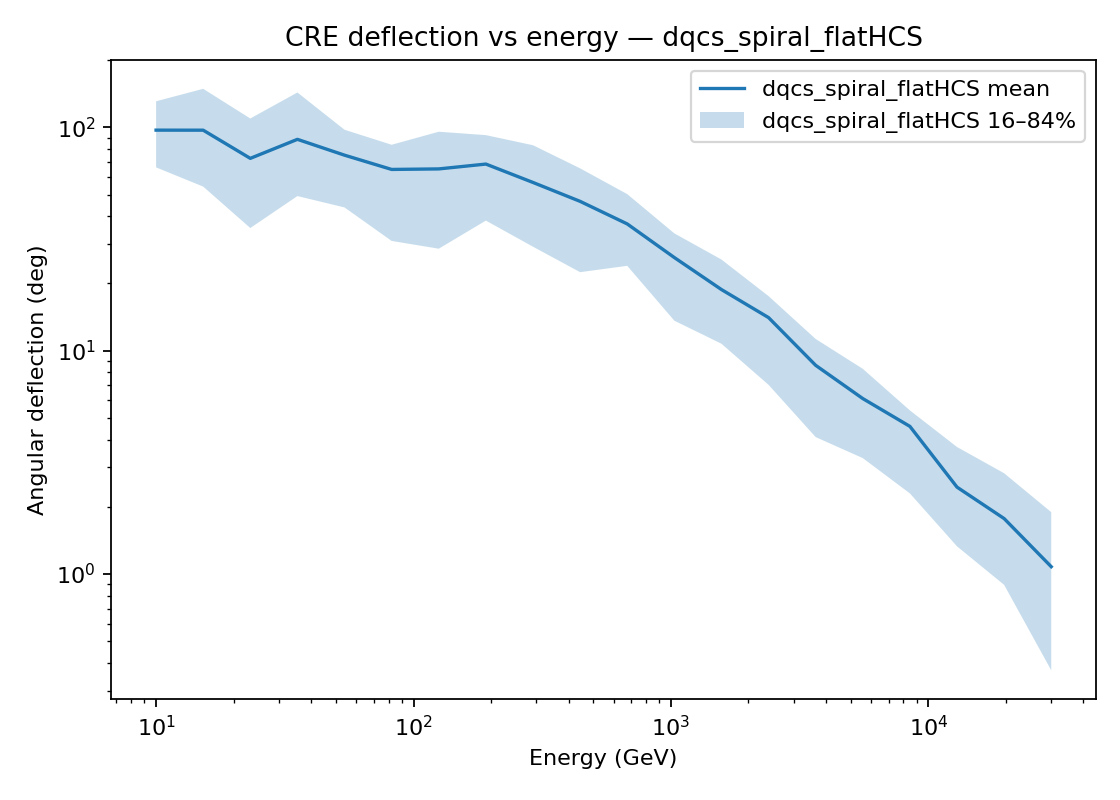}
    \caption{DQCS\,+\,spiral with \emph{flat} HCS (sector by ${\rm sign}(z)$).}
    \label{fig:dqcs_spiral_flatHCS}
  \end{subfigure}\\[0.6em]
  \begin{subfigure}[t]{0.49\textwidth}
    \centering
    \includegraphics[width=\linewidth]{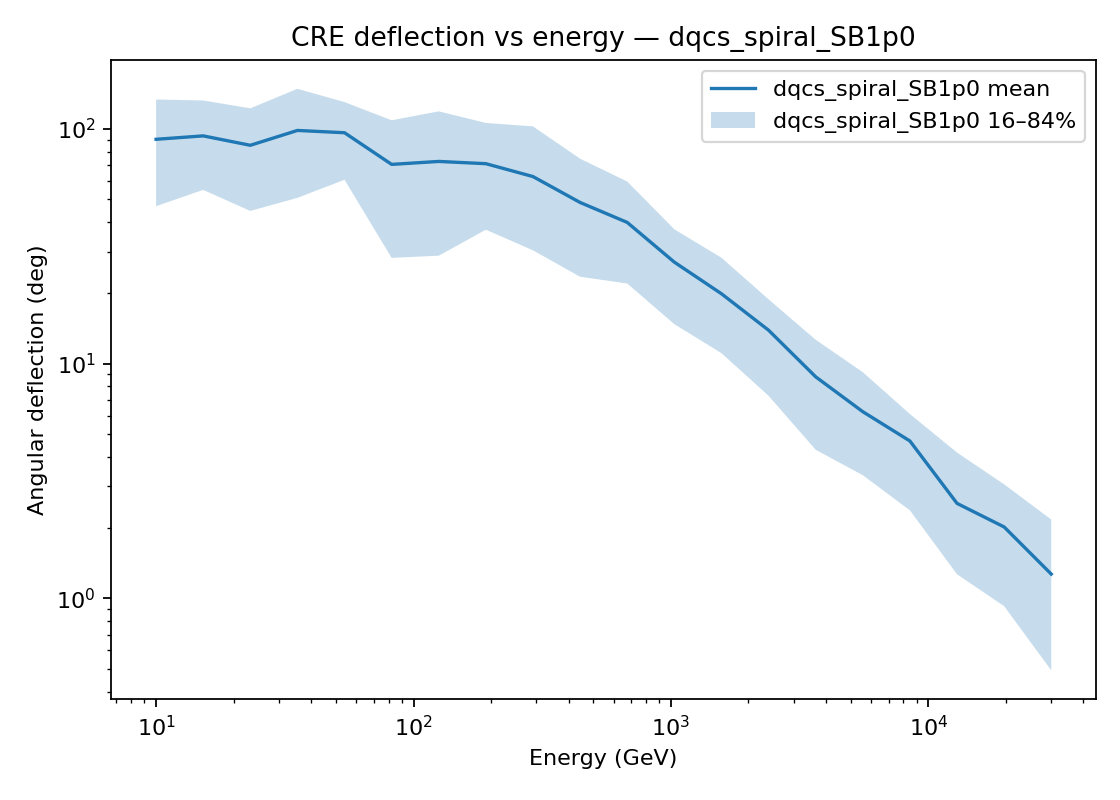}
    \caption{DQCS\,+\,spiral with Smith--Bieber ($k=1.0$).}
    \label{fig:models_SB}
  \end{subfigure}\hfill
  \begin{subfigure}[t]{0.49\textwidth}
    \centering
    \includegraphics[width=\linewidth]{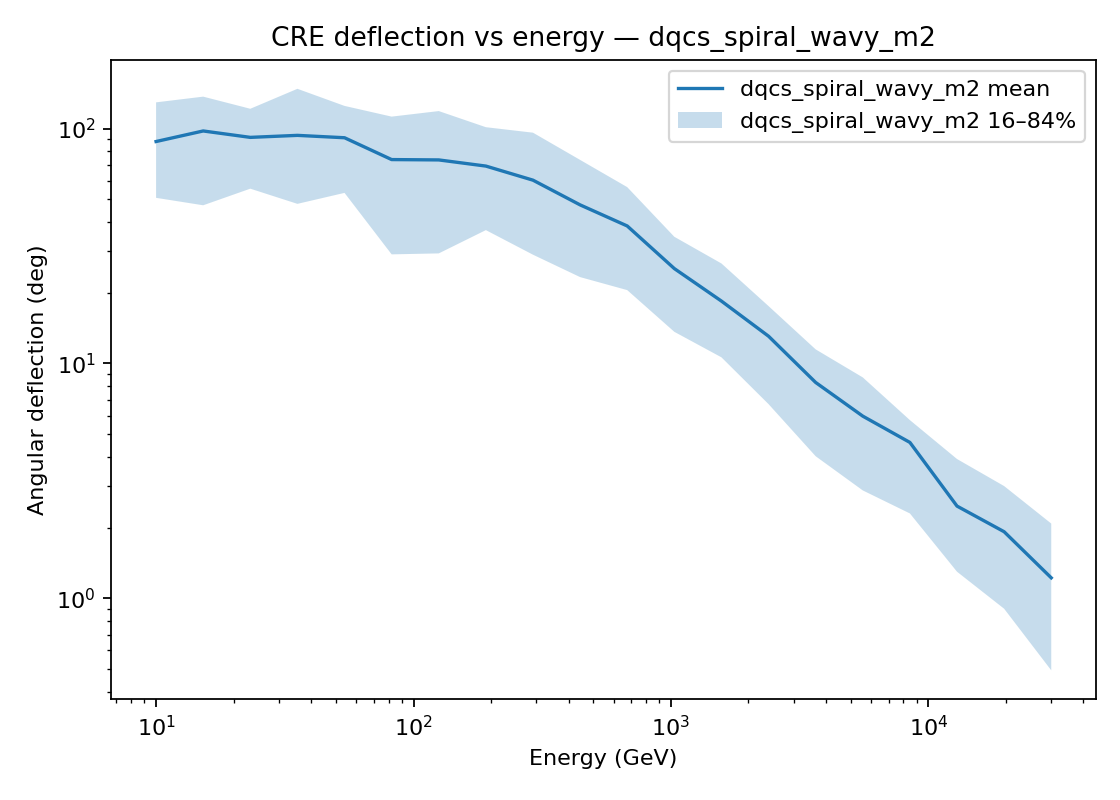}
    \caption{DQCS\,+\,spiral with wavy HCS ($m=2$).}
    \label{fig:models_wavy}
  \end{subfigure}

  \caption{{Mean heliospheric deflection and 16--84\% ranges} for six representative configurations. All runs use identical numerics (100 sky directions per energy, $r_{\rm stop}=50$\,AU, steps-per-gyro $=80$) and the same normalization of $\langle|B_r(1\,\AU)|\rangle$ and $V_{\rm sw}$.}
  \label{fig:model_comparisons_bands}
\end{figure*}

\begin{figure*}[t]
  \centering
  \includegraphics[width=0.8\linewidth]{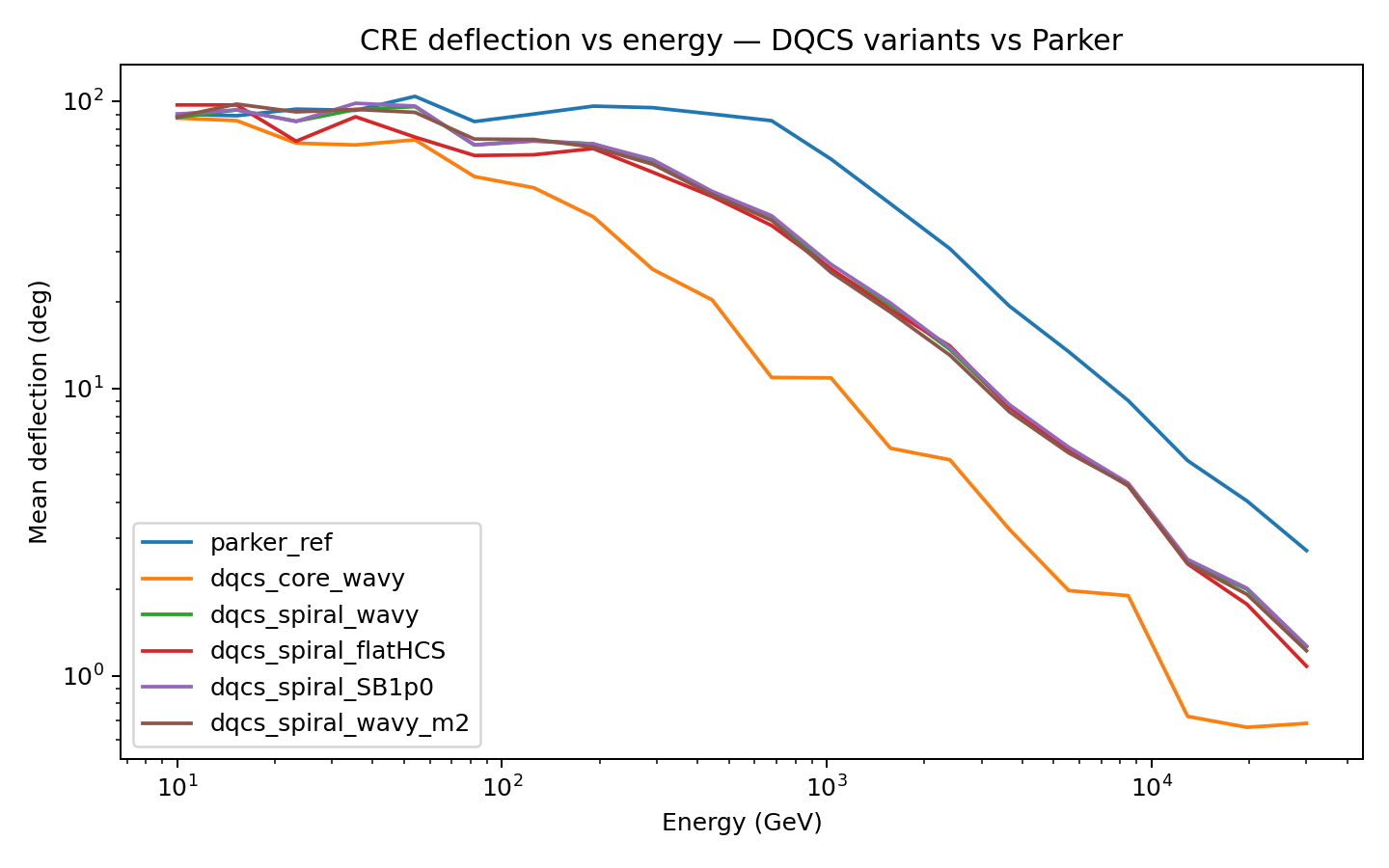}
  \caption{\textbf{Overlay of mean deflection curves} for the Parker baseline, DQCS(core), and the DQCS\,+\,spiral family (flat HCS, Smith--Bieber, and wavy HCS variants including $m=2$). Same normalization and numerics as in Fig.~\ref{fig:model_comparisons_bands}.}
  \label{fig:model_comparisons_overlay}
\end{figure*}

\subsection{Model Comparisons}
\label{sec:model_comparisons}

Figure~\ref{fig:model_comparisons_bands} contrasts a pure Parker spiral with increasingly realistic large–scale topologies based on a dipole\,+\,quadrupole\,+\,current–sheet (DQCS) coronal field. Because all panels share the same $\langle|B_r(1\,\AU)|\rangle$ and $V_{\rm sw}$ normalization, differences trace geometry and azimuthal winding rather than trivial rescaling.

{\color{black}The observed trend of decreasing deflection with increasing particle energy can be directly understood in terms of the corresponding Larmor radius, $r_{\rm L}=E/(eB)$.  
For a nominal field strength of $B\simeq5$~nT at 1~AU, the Larmor radius spans from $\sim10^{-3}$~AU at 10~GeV to $\sim10$~AU at 1~TeV.  
The transition from strong to weak deflection thus occurs when $r_{\rm L}$ becomes comparable to the characteristic heliospheric scale ($\sim1$–10~AU), in agreement with the plateau region discussed above.  
For models with larger or smaller field normalizations, this transition shifts approximately inversely with $B$, as confirmed by our tests in Section~3.}

\paragraph{Parker vs.\ DQCS(core) (panels~\ref{fig:models_parker}–\ref{fig:dqcs_core_wavy}).}
Relative to Parker’s tightly wound azimuthal structure, the DQCS core redistributes open flux and relaxes $B_\phi$ at 1\,AU. With no explicit spiral term, many lines of sight experience reduced path–integrated curvature, yielding smaller means and narrower bands above a few$\times10^2$\,GeV. This trend strengthens into the TeV range where gyroradii become comparable to inner–heliosphere scales.

\paragraph{Adding a spiral and controlling the HCS (panels~\ref{fig:dqcs_spiral_flatHCS}, \ref{fig:dqcs_spiral_wavy}).}
Superposing a Parker–like $B_\phi$ on the DQCS core partially restores curvature. With a \emph{flat} HCS (panel~\ref{fig:dqcs_spiral_flatHCS}), mean deflections rise at sub–TeV energies relative to DQCS(core) yet remain below Parker at $\gtrsim$TeV. Allowing a tilted, wavy HCS (panel~\ref{fig:dqcs_spiral_wavy}) leaves the mean close to the flat–HCS case but \emph{broadens} the 16--84\% band below $\sim$TeV, reflecting enhanced sky–to–sky variability from sector crossings and latitude–dependent access paths.

\paragraph{Smith--Bieber and higher–order HCS structure (panels~\ref{fig:models_SB}, \ref{fig:models_wavy}).}
Strengthening the azimuthal winding via Smith--Bieber ($k=1.0$; panel~\ref{fig:models_SB}) modestly boosts sub–TeV deflections relative to the flat–HCS DQCS\,+\,spiral case, consistent with tighter local curvature; the difference fades at multi–TeV where $r_L$ dominates. Increasing the HCS wave number to $m=2$ (panel~\ref{fig:models_wavy}) primarily increases the dispersion at $E\!\sim\!0.1$–$1$\,TeV without a large shift in the mean, as multiple sheet encounters diversify individual path histories.

\paragraph{Summary and interpretation.}
Across all DQCS–based variants, large–scale bending is typically \emph{smaller} than in a pure Parker spiral above $\sim$TeV, while sub–TeV behavior is controlled by how much azimuthal winding is reintroduced ($B_\phi$ via Parker/Smith--Bieber) and how frequently trajectories traverse the HCS (flat vs.\ wavy, $m=1$ vs.\ $m=2$). The percentile–band width tracks this variability: smoother fields (Parker or flat–HCS) yield narrower bands at high energy, whereas wavy sheets inflate the spread at lower energies. These trends are consistent with deflection scaling with the integrated curvature along the portion of the orbit that samples strong $B_\phi$ and with the incidence of sector crossings.\\

Note that the suppressed angular deflections in the \emph{DQCS core + wavy HCS} case arise from geometry rather than numerics. First, with the DQCS core the open magnetic flux is redistributed toward higher latitudes and the field near the ecliptic is more nearly radial than in Parker–like or spiral–augmented variants; at fixed normalization of $|B_r|(1~\mathrm{AU})$ this reduces the azimuthal component $B_\phi$ and thus the curvature of guiding centers that sets the dominant $v\times B$ rotation. Equivalently, the local field–line curvature radius is larger and the path–integrated $B_\perp$ through the inner tens of AU is smaller, yielding a lower mean deflection at a given rigidity. Second, although the HCS is wavy, the sheet in the core model is thin and the sector pattern alternates with relatively small $B_\phi$ leverage, so successive sheet encounters tend to \emph{partially cancel} accumulated bending rather than amplify it; this broadens the distribution at low energy without raising the mean. Finally, many launch directions in the core topology access higher latitudes earlier (where the spiral winding is intrinsically weaker), further shortening the effective lever arm for azimuthal turning. Taken together—more radial field near the ecliptic, weaker $B_\phi$, larger curvature radii, and partial cancellation across alternating sectors—these features explain why the ``core, wavy'' curves sit systematically below the Parker, DQCS+spiral (flat), and DQCS+spiral (wavy) cases across the sub–TeV to TeV range.

For practical anisotropy work, the differences in Fig.~\ref{fig:model_comparisons_overlay} translate into model-dependent threshold energies \(E_{\rm crit}(\theta_{\rm inst})\): relative to Parker, DQCS–based models raise $E_{\rm crit}$ at low energies only slightly (due to added $B_\phi$), but \emph{lower} $E_{\rm crit}$ in the TeV range where the mean deflection is a factor of a few smaller. Consequently, conclusions about CRE dipoles near a degree should quote a systematic envelope that spans these model families.

\emph{Implications for the positron excess.}
The energy–dependent bending illustrated in Fig.~\ref{fig:model_comparisons_overlay} is directly relevant to directional tests of the PAMELA/AMS positron excess \citep{Adriani2009Nature,Aguilar2013PRL,Aguilar2019PRL}. In source scenarios where one (or a few) nearby pulsars dominate the local positron budget—most prominently Geminga and Monogem \citep{Hooper2009JCAP,Yuksel2009PRL,LindenProfumo2013ApJ,Abeysekara2017Science}—the interstellar arrival distribution at the heliopause can carry a broad hemispheric excess pointing toward the candidate. Our calculations show that, once this interstellar pattern is propagated through the heliosphere, a \emph{net hemispheric asymmetry in the positron sky can survive at high energies} (e.g., $\gtrsim$~TeV) for field configurations with stronger azimuthal winding and modest HCS tilts. However, the same configurations induce tens–of–degree deflections at sub–TeV energies; in that regime the sheet geometry and charge–sign drifts generically wash out the large–scale gradient, diluting any hemispheric imprint from a single nearby pulsar. Consequently, directional tests of the pulsar origin of the excess are most informative at the top of the AMS energy range and above, while at a few $\times 10^2$~GeV and below the heliospheric leg must be folded in as a foreground that can rotate and smooth the expected gradient. This interpretation is consistent with the small upper limits on large–scale CRE anisotropy (see e.g. Ref.~\cite{Ackermann2010FermiCREAniso}) and with the picture emerging from TeV–halo observations, which establish strong local injection from Geminga/Monogem but also suggest transport conditions that further blur directional information before particles reach the heliosphere \citep{Abeysekara2017Science}.
{\color{black}At TeV energies, electrons and positrons experience opposite Lorentz forces in the large-scale heliospheric field, leading to mirror-image deflection patterns.  
For the Parker field configuration, this charge-sign dependence produces an angular separation of order $\Delta\theta_{\rm e^{\pm}}\!\sim\!2\,r_{\rm L}^{-1}\!\!\int B_{\perp}\,ds$, corresponding to $\sim1$–$2^{\circ}$ at 1~TeV.  
Such differences imply that the two charge populations may probe slightly distinct entry regions at the heliopause, potentially diluting any intrinsic anisotropy.  
This effect is smaller than the angular resolution of current detectors but could become relevant for next-generation wide-field instruments or for stacked multi-year anisotropy analyses.}


\begin{figure*}[!h]
  \centering
  \begin{subfigure}[t]{0.49\textwidth}
    \centering
    \includegraphics[width=\linewidth]{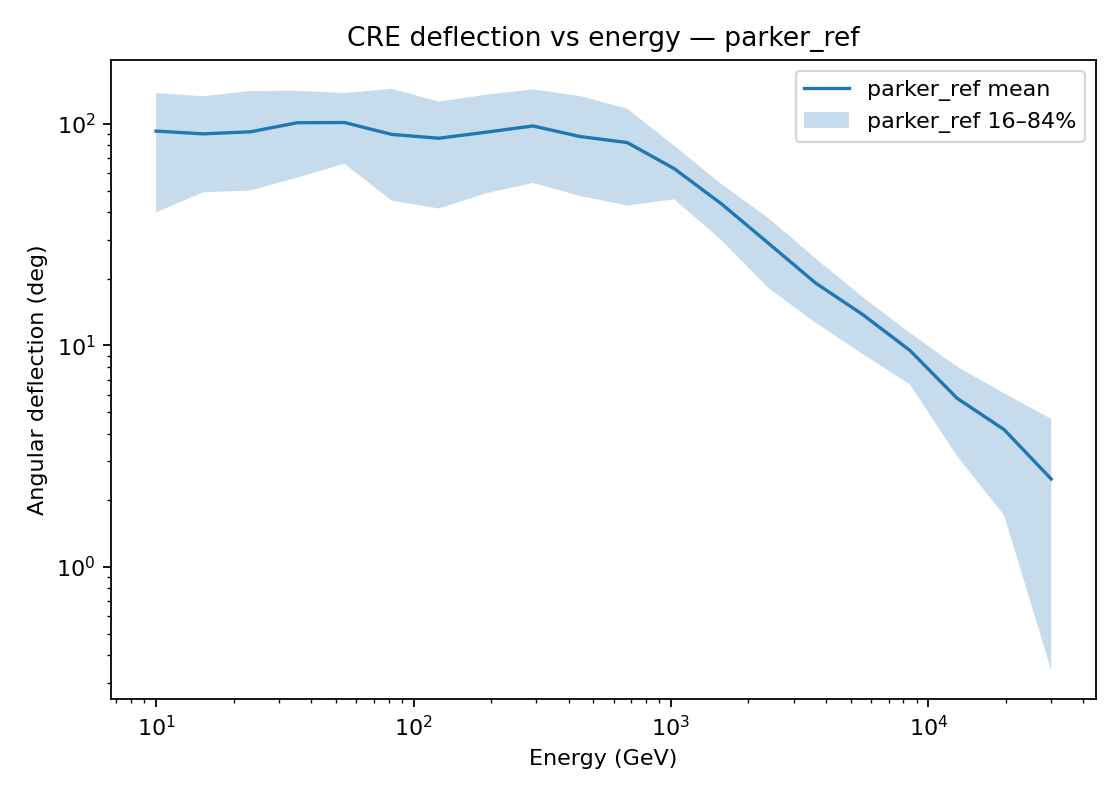}
    \caption{Parker reference (unipolar).}
    \label{fig:hcs_parker}
  \end{subfigure}\hfill
  \begin{subfigure}[t]{0.49\textwidth}
    \centering
    \includegraphics[width=\linewidth]{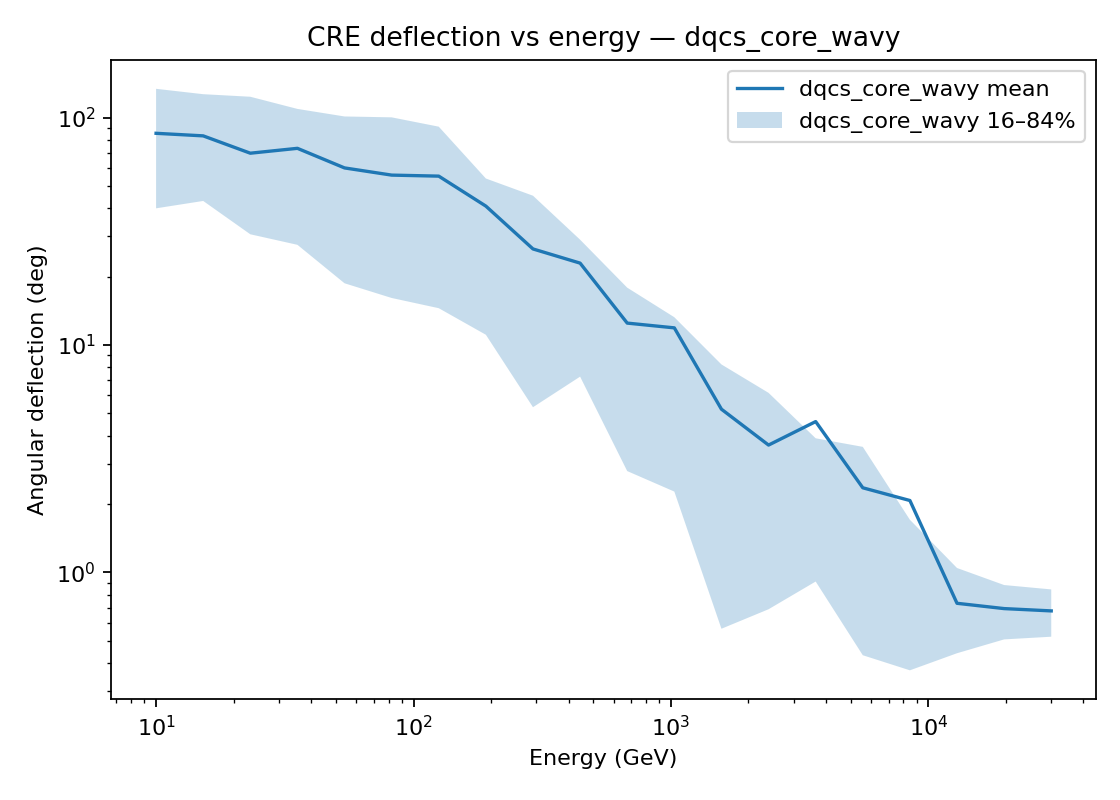}
    \caption{DQCS core with wavy HCS ($m=1$).}
    \label{fig:hcs_core_m1}
  \end{subfigure}\\[0.6em]
  \begin{subfigure}[t]{0.49\textwidth}
    \centering
    \includegraphics[width=\linewidth]{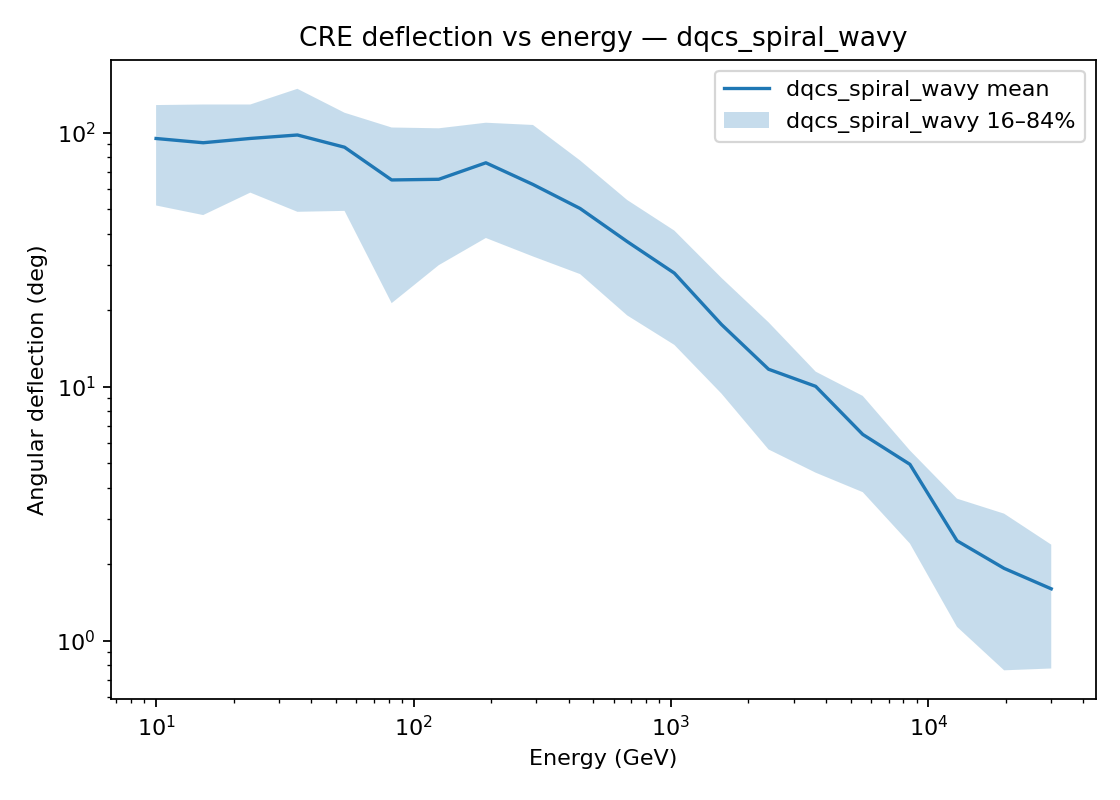}
    \caption{DQCS + spiral, wavy HCS ($m=1$).}
    \label{fig:hcs_spiral_m1}
  \end{subfigure}\hfill
  \begin{subfigure}[t]{0.49\textwidth}
    \centering
    \includegraphics[width=\linewidth]{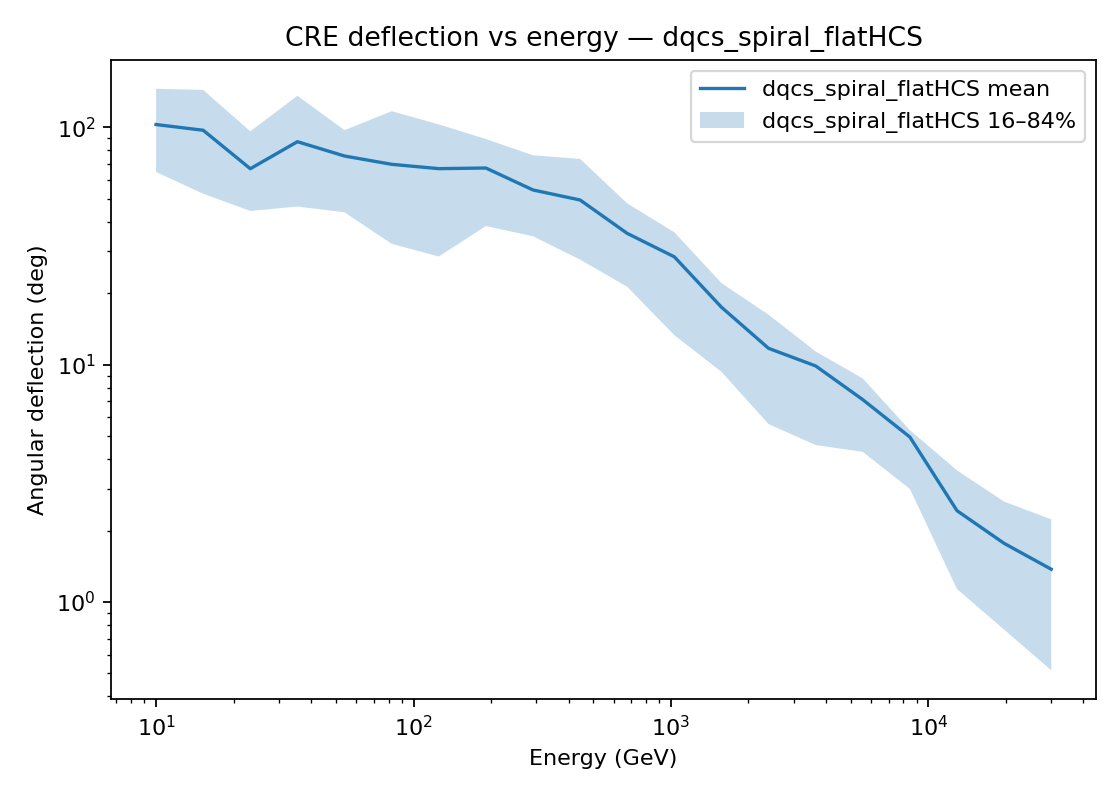}
    \caption{DQCS + spiral with flat HCS (sector by sign($z$)).}
    \label{fig:hcs_spiral_flat}
  \end{subfigure}\\[0.6em]
  \begin{subfigure}[t]{0.49\textwidth}
    \centering
    \includegraphics[width=\linewidth]{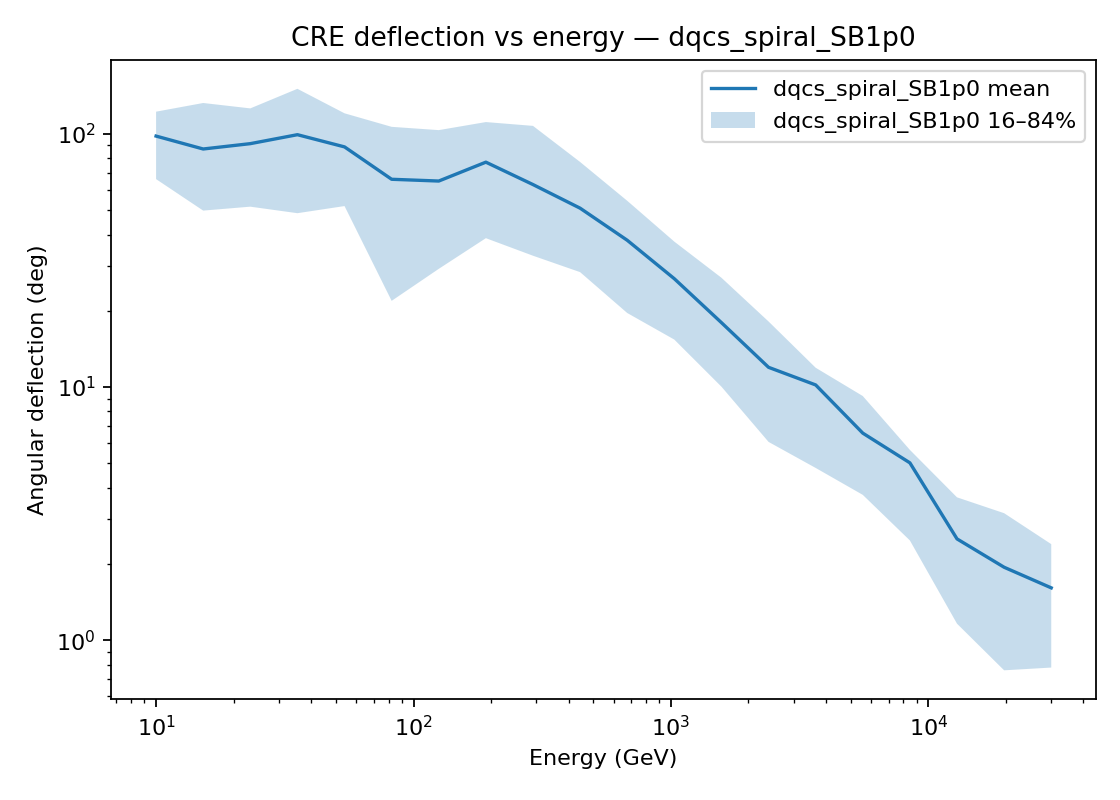}
    \caption{DQCS + spiral, stronger Smith--Bieber.}
    \label{fig:hcs_spiral_sb}
  \end{subfigure}\hfill
  \begin{subfigure}[t]{0.49\textwidth}
    \centering
    \includegraphics[width=\linewidth]{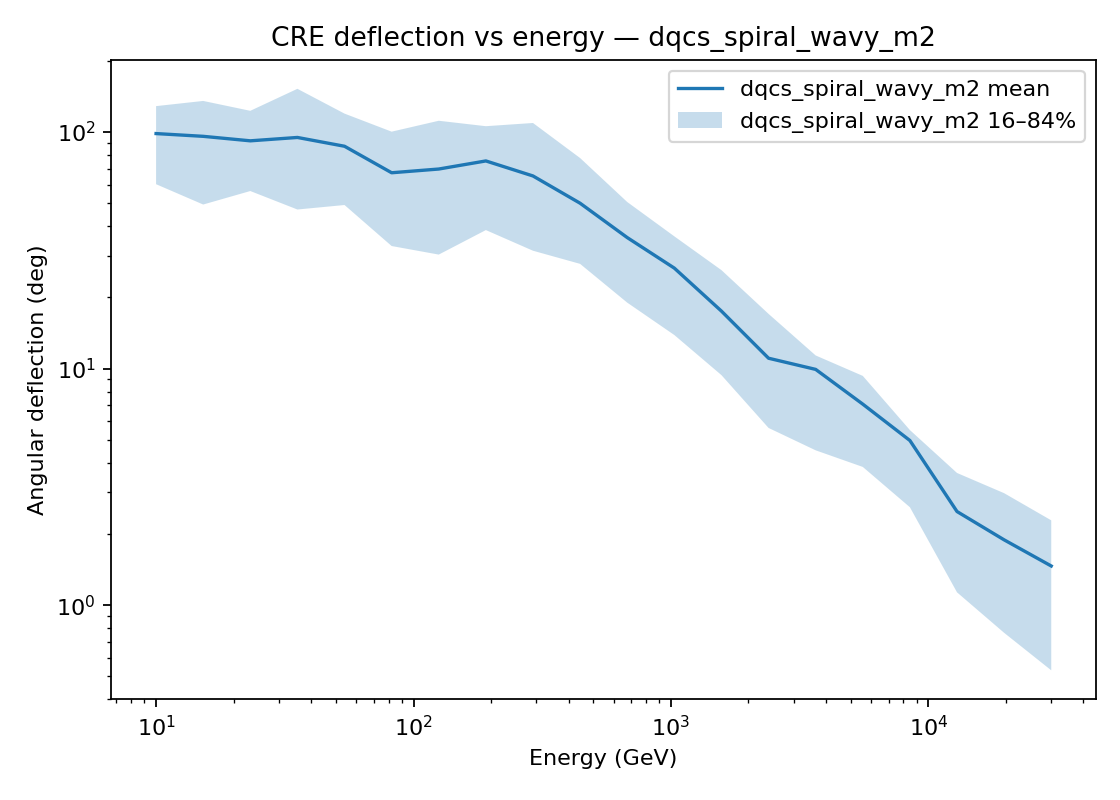}
    \caption{DQCS + spiral, wavy HCS ($m=2$).}
    \label{fig:hcs_spiral_m2}
  \end{subfigure}

  \caption{\textbf{HCS geometry and deflection bands.}
  Mean heliospheric deflection and 16--84\% ranges for a Parker reference and DQCS-based configurations with tilted/wavy heliospheric current sheet (HCS). Shared numerics across panels: 100 sky directions per energy, $r_{\rm stop}=50$\,AU, steps-per-gyro $=80$, and common $\langle|B_r(1\,\AU)|\rangle$ and $V_{\rm sw}$.}
  \label{fig:hcs_panels}
\end{figure*}

\begin{figure*}[t]
  \centering
  \includegraphics[width=0.78\linewidth]{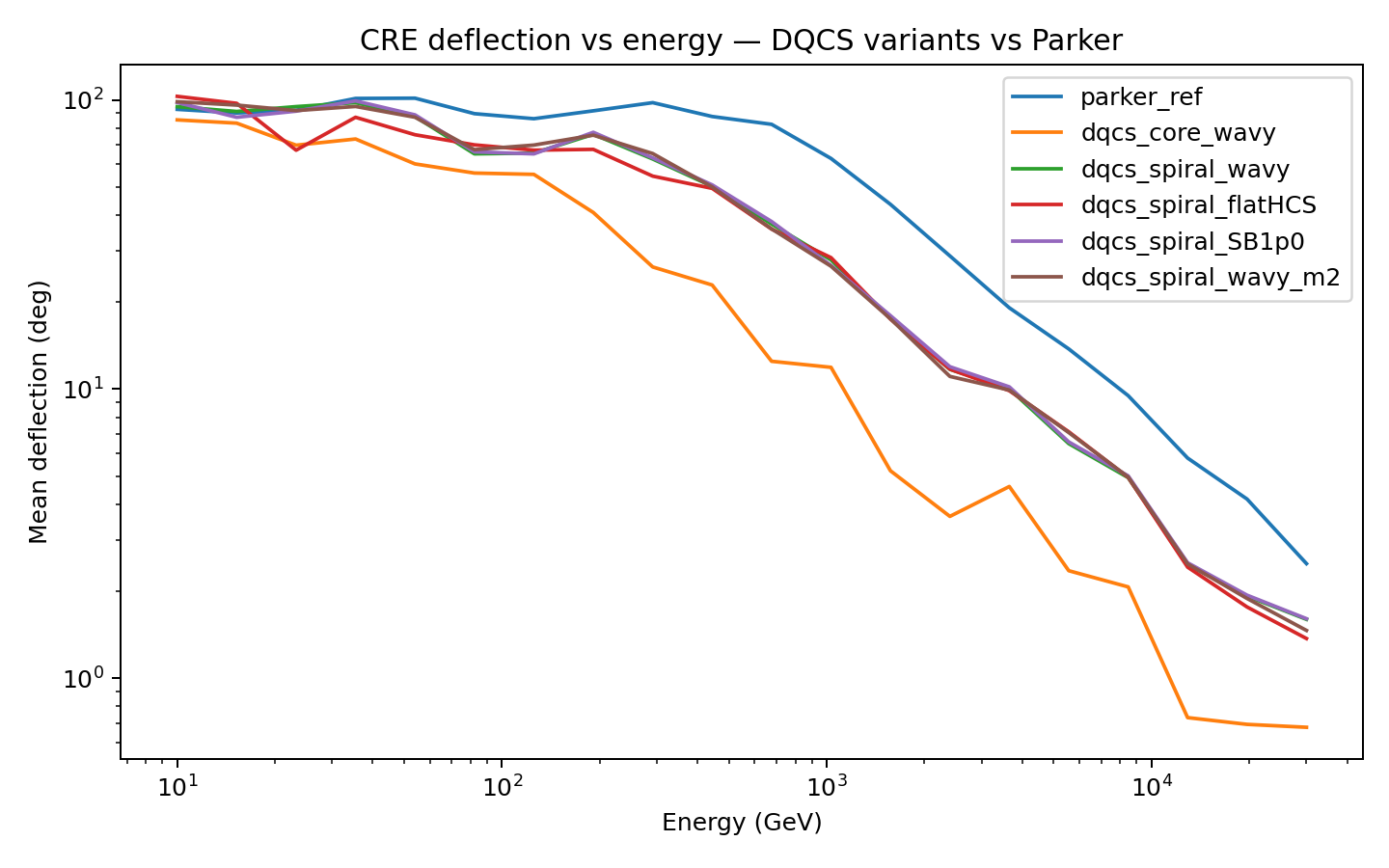}
  \caption{\textbf{Mean-curve overlay for HCS cases.}
  Overlay of mean deflection curves for the $m=1$ cases in Fig.~\ref{fig:hcs_panels} (panels~\subref{fig:hcs_parker}--\subref{fig:hcs_spiral_sb}); the $m=2$ curve from panel~\subref{fig:hcs_spiral_m2} can be added for comparison where noted. Same normalization and numerics as Fig.~\ref{fig:hcs_panels}.}
  \label{fig:hcs_overlay}
\end{figure*}

\subsection{HCS Geometry: Tilt and Waviness}
\label{sec:hcs_geometry}

The preceding \emph{Model Comparisons} section varied the \emph{global field topology} (Parker vs.\ DQCS, with/without additional azimuthal winding and Smith--Bieber enhancement) and showed that those choices chiefly set the overall \emph{level} of bending—Parker generally producing the largest mean deflections, while DQCS-based models reduce bending above $\sim$TeV by relaxing azimuthal winding.  
In this section we hold that global topology fixed (within the DQCS family, plus a Parker reference) and isolate the role of the \emph{heliospheric current sheet} (HCS): its tilt, waviness, and sector structure. The key difference in outcomes is that HCS geometry predominantly modulates the \emph{sky-to-sky variance} at sub-TeV energies, with comparatively modest shifts in the mean, whereas topology changes in Sec.~\ref{sec:model_comparisons} reshaped the \emph{mean curves} themselves.

\paragraph{Core vs.\ spiral with a wavy HCS.}
Panels~\subref{fig:hcs_core_m1} and \subref{fig:hcs_spiral_m1} compare the DQCS core (no local Parker-like $B_\phi$) to DQCS+spiral, both with a wavy $m=1$ HCS. Adding $B_\phi$ increases the path-integrated curvature and lifts the mean deflection below the TeV scale, but both cases converge toward similar means at multi-TeV where the Larmor radius dominates. This \emph{mean} shift is smaller than the Parker$\rightarrow$DQCS change seen earlier, underscoring that HCS geometry mainly alters \emph{dispersion}, not the asymptotic scaling. {\color{black}Notice that the distributions of deflection angles at fixed energy are mildly asymmetric, especially in the transition region between 0.3 and 3~TeV where a small subset of trajectories experiences larger-than-average bending due to the crossing of the heliospheric current sheet.  
The solid line (mean deflection) may therefore appear offset from the center of the dispersion band, which reflects the 68\% central interval of the distribution rather than the symmetric $\pm1\sigma$ range.  
At both lower and higher energies, the distributions become approximately Gaussian, and the mean lies well within the dispersion envelope.}

\paragraph{Flat vs.\ wavy HCS.}
Comparing panels~\subref{fig:hcs_spiral_flat} and \subref{fig:hcs_spiral_m1} shows that replacing a flat sector boundary with a tilted, wavy sheet broadens the 16--84\% bands—especially at $E\!\sim\!30$--$300$\,GeV—while leaving the mean only modestly affected. Physically, additional undulations increase the chance and multiplicity of sector crossings and drift-rich path segments; small changes in launch direction translate into larger path-length and curvature differences across the sky.

\paragraph{Smith--Bieber and field winding.}
Strengthening the local azimuthal field (panel~\subref{fig:hcs_spiral_sb}) modestly raises sub-TeV means relative to the flat-HCS DQCS+spiral case, consistent with tighter winding, but the difference is again minor compared to the Parker vs.\ DQCS split from Sec.~\ref{sec:model_comparisons}. At TeV energies the mean curves nearly coincide.

\paragraph{Higher waviness ($m=2$).}
Introducing a second wave in longitude (panel~\ref{fig:hcs_spiral_m2}) further amplifies the percentile width at low energies while keeping the mean close to the $m=1$ case. The mean-overlay in Fig.~\ref{fig:hcs_overlay} highlights that ordering: Parker remains largest, DQCS(core) is the most suppressed, and the DQCS+spiral variants (flat/wavy, SB) cluster together in mean but differ in spread.

\paragraph{Summary of contrasts.}
Relative to Sec.~\ref{sec:model_comparisons}, where global topology controlled the \emph{mean} deflection level and its TeV asymptote, the HCS-focused variations here primarily control the \emph{direction-to-direction scatter} at sub-TeV energies through sector crossings and charge-sign drifts, with only secondary impact on the mean curve. Consequently, anisotropy forecasts should (i) budget larger systematic envelopes in energy ranges where the line of sight is likely to intersect a tilted/wavy HCS, and (ii) expect similar TeV-scale means across HCS variants once the gyroradius exceeds the HCS curvature scale.


\subsection{Solar-Cycle Sensitivity: \texorpdfstring{$B_r(1\,\AU)$}{Br(1 AU)} and \texorpdfstring{$V_{\rm sw}$}{Vsw}}
\label{sec:cycle}

\begin{figure*}[t]
  \centering
  \begin{subfigure}[t]{0.49\textwidth}
    \centering
    \includegraphics[width=\linewidth]{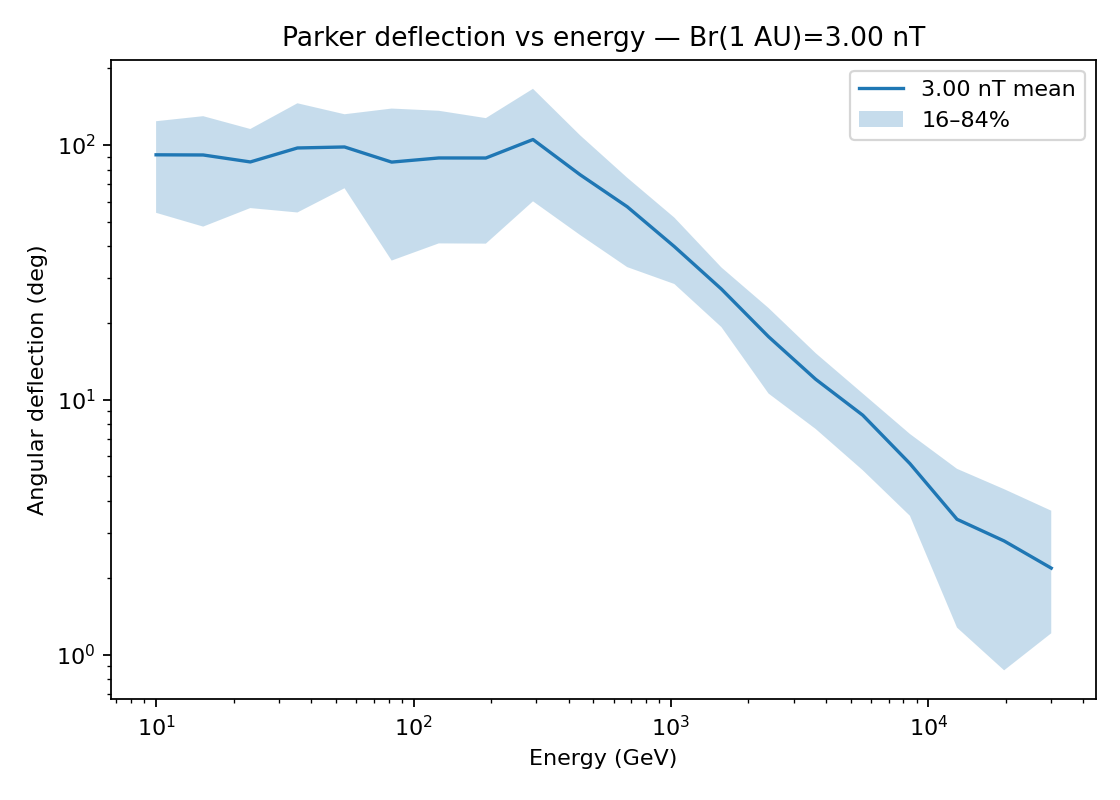}
    \caption{Parker, $B_r(1\,\AU)=3$ nT.}
  \end{subfigure}\hfill
  \begin{subfigure}[t]{0.49\textwidth}
    \centering
    \includegraphics[width=\linewidth]{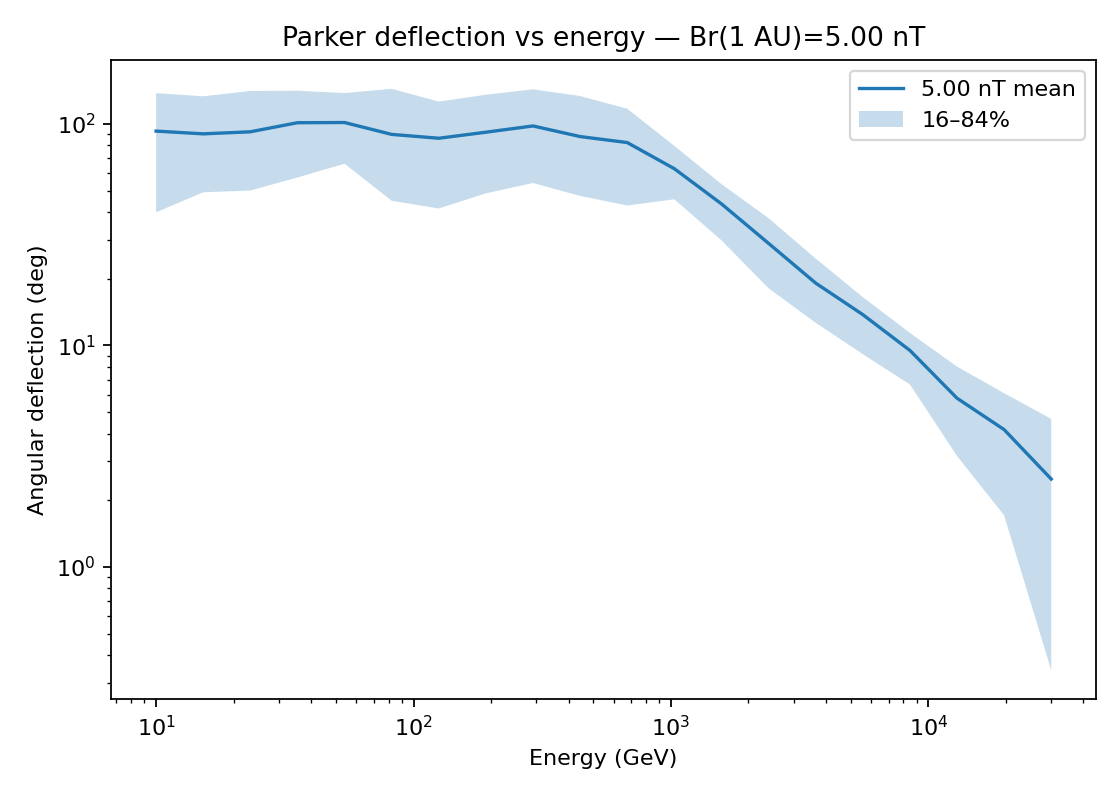}
    \caption{Parker, $B_r(1\,\AU)=5$ nT.}
  \end{subfigure}\\[0.6em]
  \begin{subfigure}[t]{0.49\textwidth}
    \centering
    \includegraphics[width=\linewidth]{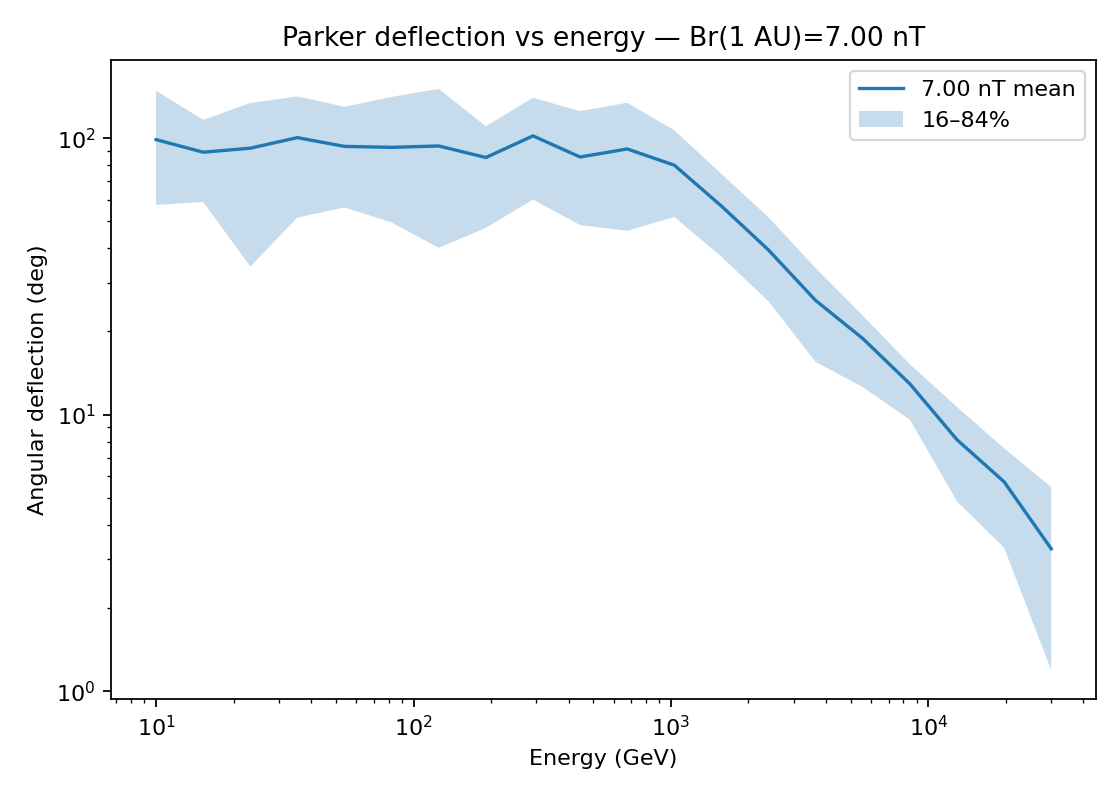}
    \caption{Parker, $B_r(1\,\AU)=7$ nT.}
  \end{subfigure}\hfill
  \begin{subfigure}[t]{0.49\textwidth}
    \centering
    \includegraphics[width=\linewidth]{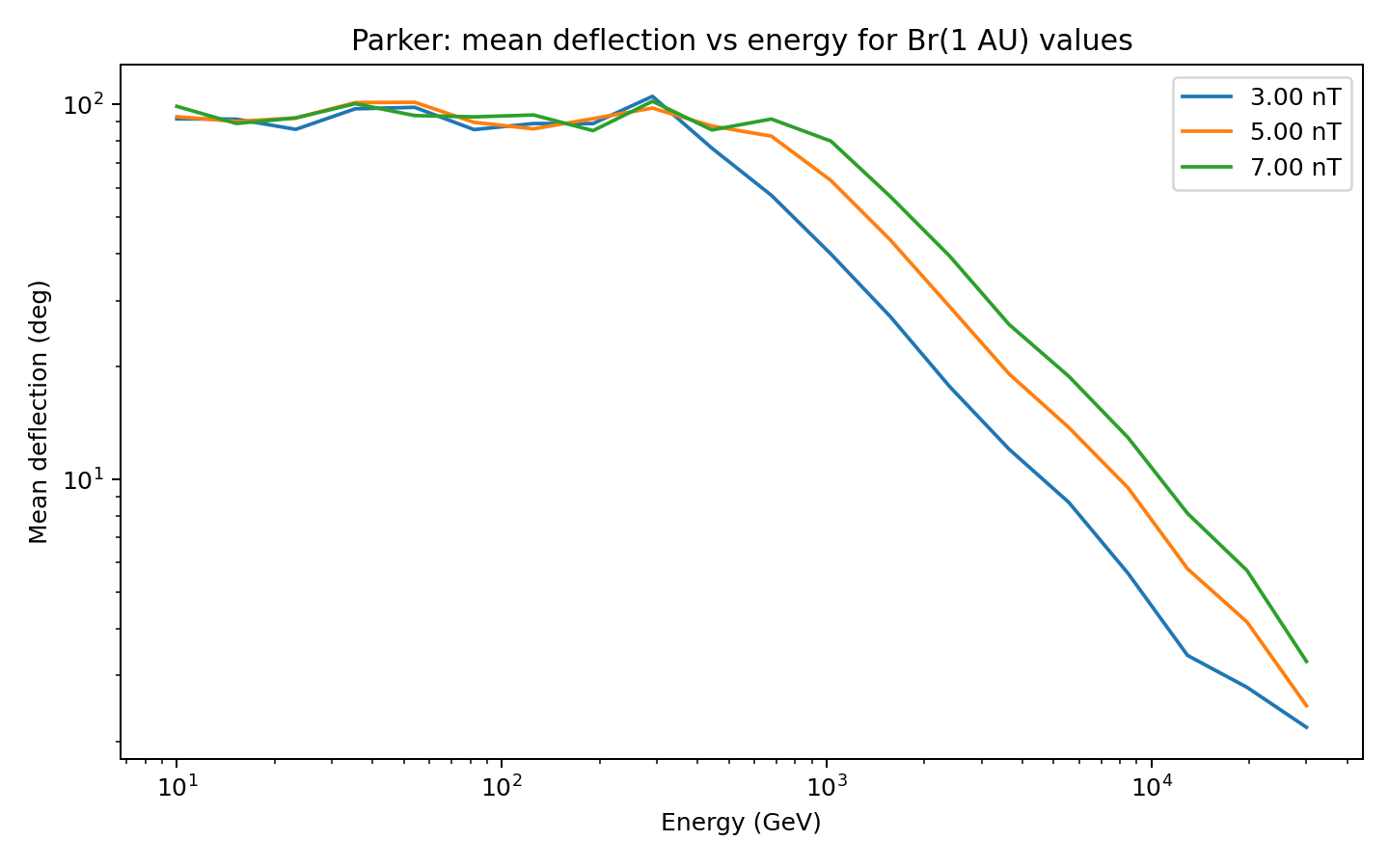}
    \caption{Overlay of means for 3/5/7 nT.}
  \end{subfigure}
  \caption{Mean (solid) and 16--84\% bands (shaded) for three field strengths representative of solar-min to active conditions; right panel overlays the means. All runs share identical numerics and sky sampling.}
  \label{fig:fig5a_parkerB}
\end{figure*}

A central solar-cycle lever on heliospheric bending is the large-scale field strength at 1\,AU, $B_r(1\,\AU)$, which scales the Parker spiral and any spiral-like component added to more complex models. To leading order, the characteristic deflection angle scales as
\begin{equation}
\theta_{\rm defl}\ \propto\ \frac{Z e}{p}\, \int \! |\mathbf{B}_\perp|\,\mathrm{d}\ell
\ \sim\ \frac{B_{\rm eff}}{E}\,,
\end{equation}
so increasing $B_r(1\,\AU)$ shifts the entire deflection--energy curve upward and moves the “negligible-bending” threshold to higher energy. Figure~\ref{fig:fig5a_parkerB} shows this explicitly for the Parker baseline: from $B_r(1\,\AU)=3\,$nT (solar-min like) to 7\,nT (active conditions), the mean deflection above $\sim$\,TeV increases by a factor of $\simeq$2--3, and the transition from $\mathcal{O}(10^{2})$\,deg to single-digit degrees migrates from $\sim$\,few\,$\times 10^3$\,GeV toward $\sim$\,few\,$\times 10^4$\,GeV. The spread (16--84\%) also broadens slightly with stronger fields, reflecting enhanced latitude dependence (larger $\Bphi$ near the equator) and more pronounced azimuthal winding.

\begin{figure*}[!h]
  \centering
  \begin{subfigure}[t]{0.49\textwidth}
    \centering
    \includegraphics[width=\linewidth]{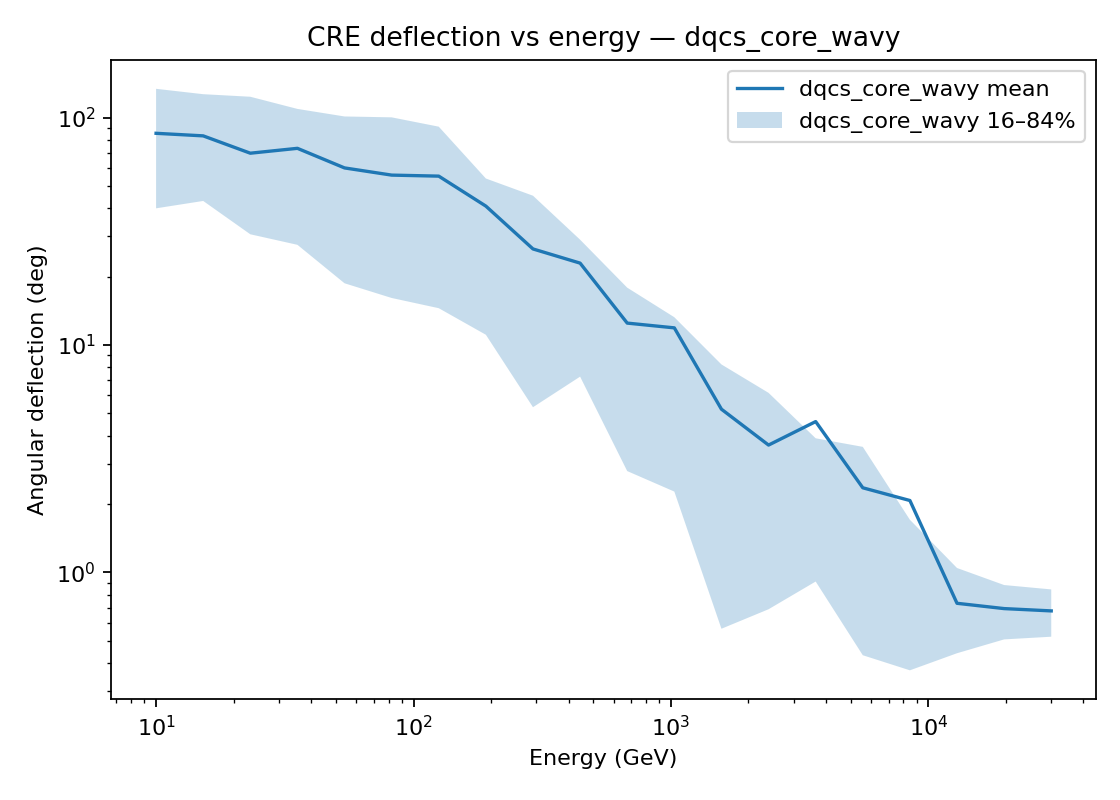}
    \caption{DQCS(core)+wavy HCS.}
  \end{subfigure}\hfill
  \begin{subfigure}[t]{0.49\textwidth}
    \centering
    \includegraphics[width=\linewidth]{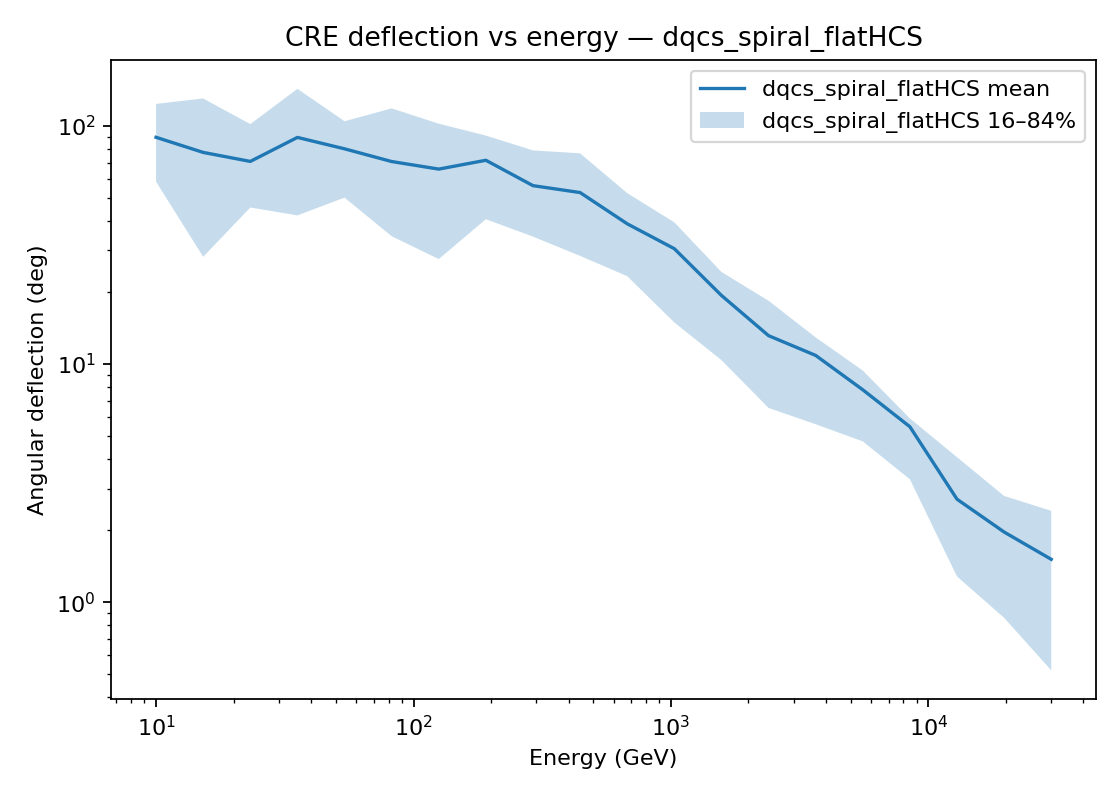}
    \caption{DQCS + spiral, flat HCS.}
  \end{subfigure}\\[0.6em]
  \begin{subfigure}[t]{0.49\textwidth}
    \centering
    \includegraphics[width=\linewidth]{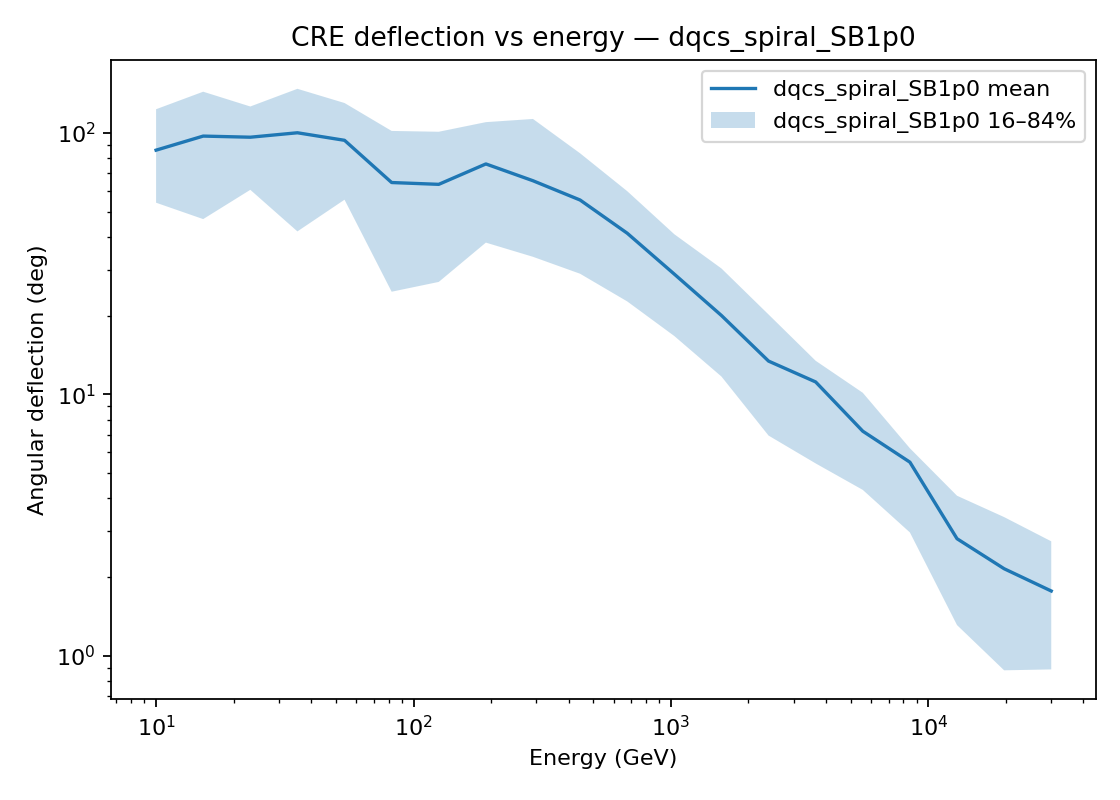}
    \caption{DQCS + spiral, Smith--Bieber.}
  \end{subfigure}\hfill
  \begin{subfigure}[t]{0.49\textwidth}
    \centering
    \includegraphics[width=\linewidth]{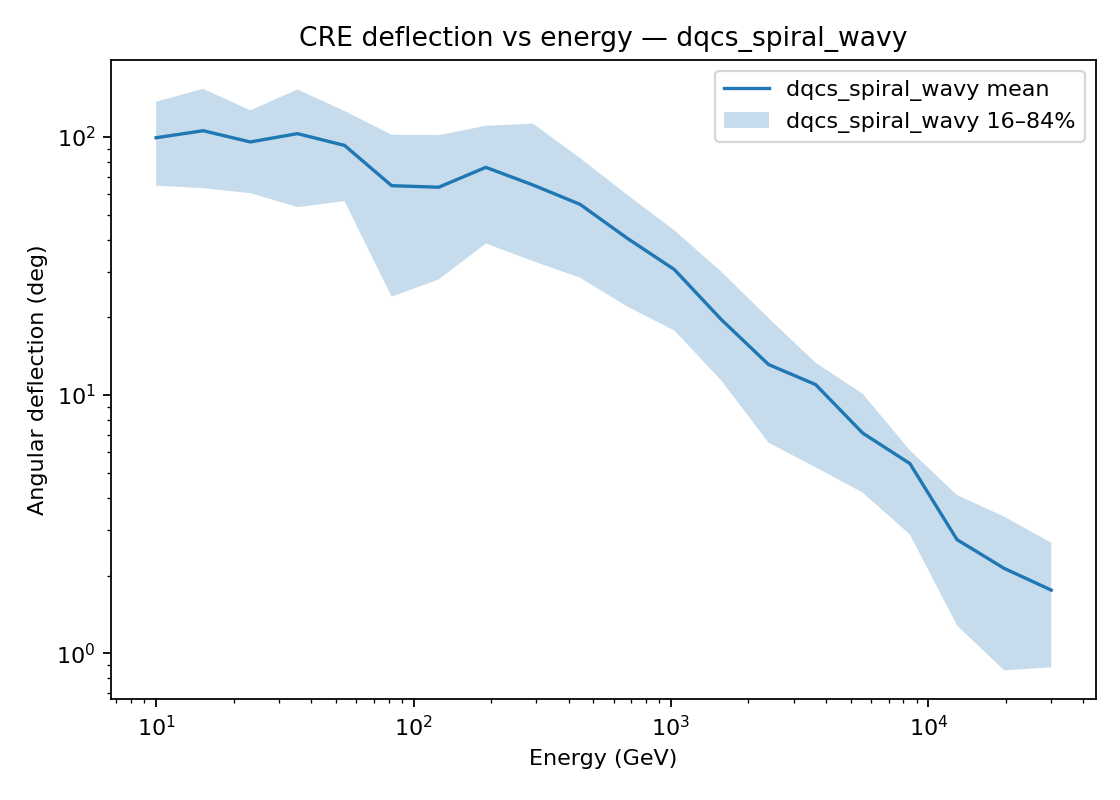}
    \caption{DQCS + spiral, wavy HCS ($m=1$).}
  \end{subfigure}\\[0.6em]
  \begin{subfigure}[t]{0.49\textwidth}
    \centering
    \includegraphics[width=\linewidth]{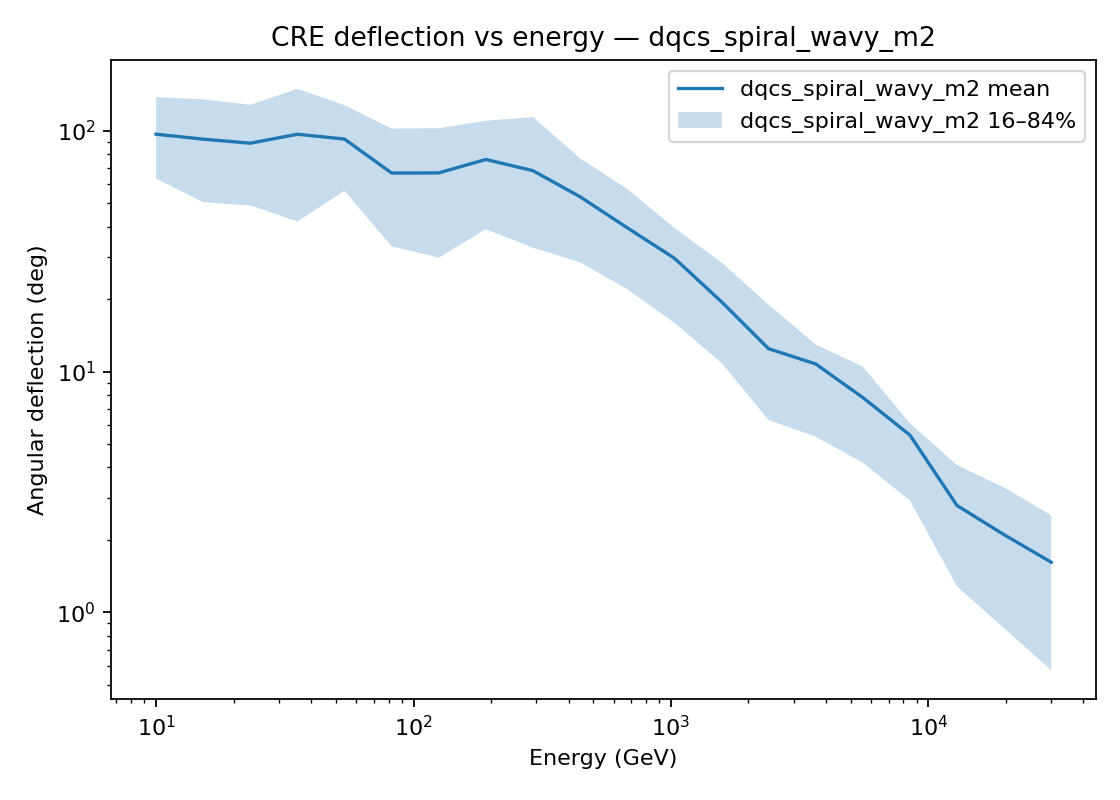}
    \caption{DQCS + spiral, wavy HCS ($m=2$).}
  \end{subfigure}\hfill
  \begin{subfigure}[t]{0.49\textwidth}
    \centering
    \includegraphics[width=\linewidth]{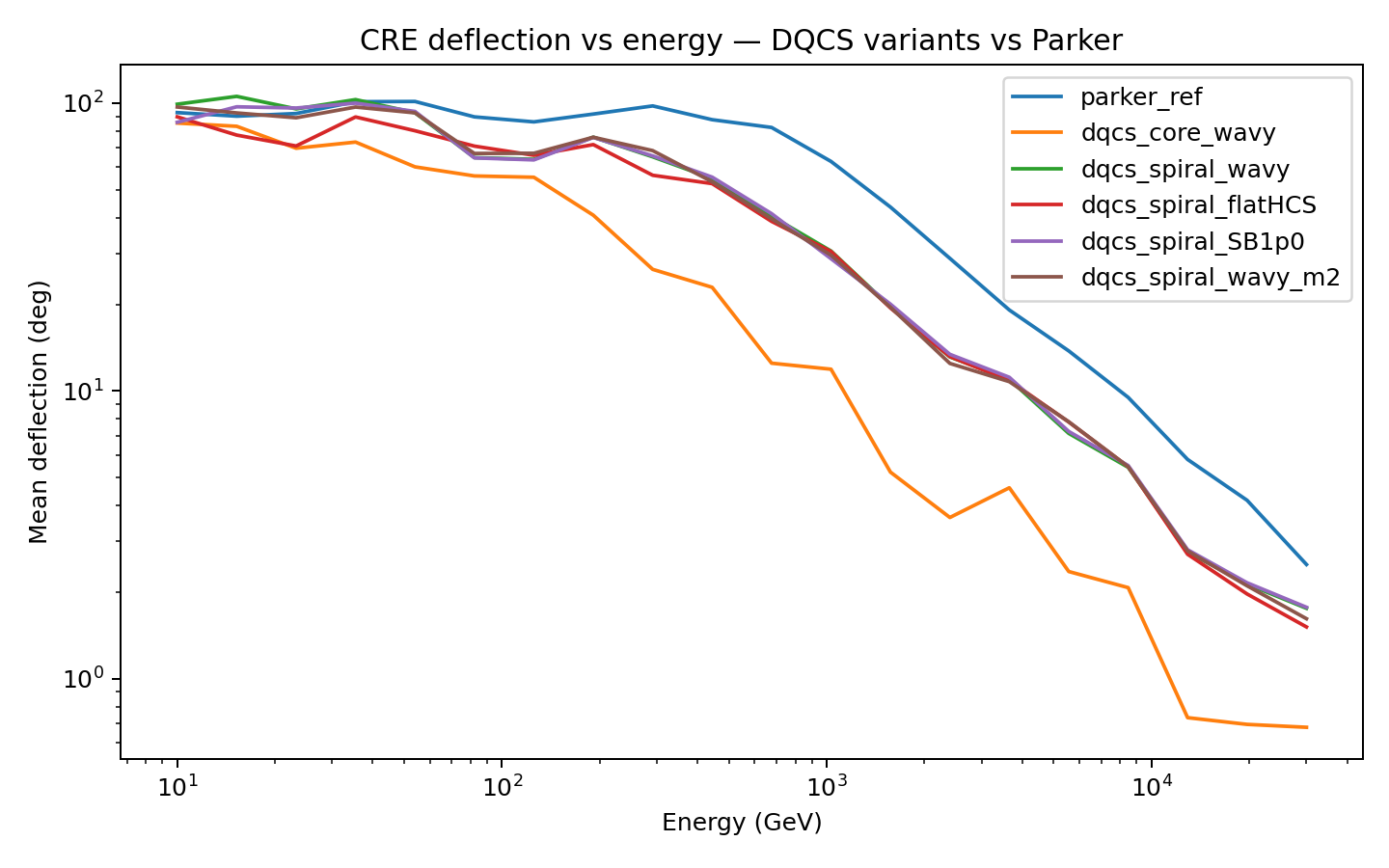}
    \caption{Overlay of means across variants.}
  \end{subfigure}
  \caption{Two-stream wind ($350/650$\,km\,s$^{-1}$ for equator/poles) reduces the effective spiral at high latitudes, shrinking deflections and narrowing percentile bands there. Smith--Bieber and wavy-HCS variants increase the spiral and/or the probability of current-sheet encounters, raising deflections and broadening the band at tens--hundreds of GeV.}
  \label{fig:fig5b_vsw}
\end{figure*}

A second lever is the solar-wind speed $V_{\rm sw}(\theta)$. In any spiral-like geometry, $\Bphi/\Br \propto (\Omega_\odot r)/V_{\rm sw}$, so faster winds reduce the azimuthal winding (hence smaller deflections), while a latitudinal contrast---slow at low latitudes and fast at high latitudes---imprints a distinct sky pattern. Figure~\ref{fig:fig5b_vsw} compares several DQCS\,+\,spiral variants under a two-stream prescription (slow $350$\,km\,s$^{-1}$ equator; fast $650$\,km\,s$^{-1}$ poles). Two robust trends emerge:
(i) models with stronger effective spiral (e.g., Smith--Bieber enhancement) produce larger deflections and a later transition to the weak-bending regime; 
(ii) introducing a latitudinal wind contrast narrows the percentile band at high energies by suppressing $\Bphi$ off the equator, while leaving low-energy behavior largely unchanged because $\Br$ dominates the integrated curvature there.
Across models, the energy at which the mean deflection drops below $\sim 5^\circ$ varies at the $\sim$\,factor of 2 level, with wavy-HCS cases tracking the flat-HCS curves at high energy but exhibiting a broader band at tens--hundreds of GeV due to current-sheet encounters and charge-sign drifts (not shown here).

{\color{black}For both the modified Parker model and the configuration including the heliospheric current sheet, the dependence of the mean deflection on the field intensity at 1~AU is qualitatively similar to that of the pure Parker field: $\langle\Delta\theta\rangle \propto B_{1{\rm AU}}^{-1}$ in the ballistic regime and approaches saturation at lower energies.  
Quantitatively, the presence of latitudinal structure or current-sheet reversals produces at most $\sim$10–15\% deviations from this scaling across the 1–10~nT range tested.  
These variations confirm that our results are robust against moderate uncertainties in the field normalization.}

\paragraph{Practical takeaway.} For instrument Point Spread Functions (PSFs) of a few tenths of a degree, the multi-TeV regime is required for Parker-like fields at $B_r(1\,\AU)\gtrsim 5$\,nT, while in fast-wind or weak-spiral conditions the same directional cleanliness can be reached at somewhat lower energy. Conversely, during active periods (higher $B_r$, slower equatorial winds) the heliospheric budget should be explicitly folded into anisotropy analyses up to tens of TeV.

\section{Discussion}
\label{sec:discussion}

Our results provide a practical, experiment–facing “transfer function’’ between large–scale heliospheric structure and the direction–reconstruction budget of high–energy CRE measurements. Across the field families explored—Parker baseline, DQCS–type topologies, spiral/Smith–Bieber enhancements, latitude–dependent winds, and tilted/wavy current sheets—the sky–averaged deflection decreases rapidly with energy, from $\mathcal{O}(10\text{–}100)^\circ$ at tens of~GeV to $\lesssim 1^\circ$ at multi–TeV. The normalization is set predominantly by \emph{geometry}: the effective path–integrated $|B_\perp|$ through the inner tens of AU and the fraction of directions that graze or cross the heliospheric current sheet (HCS).

\paragraph{What sets the negligible–bending threshold?}
A useful way to summarize the impact is via the energy $E_{\rm crit}(\theta_{\rm inst})$ at which the mean heliospheric deflection falls below a chosen angular budget (e.g., the instrument PSF, $\theta_{\rm inst}$). To first order,
\begin{equation}
\langle \Delta\theta \rangle \;\propto\; \frac{\langle |B_\perp| \rangle\, L_{\rm eff}}{E}\,,
\end{equation}
so $E_{\rm crit}$ scales linearly with the effective spiral strength and the geometric exposure to sheet encounters. {\color{black}The quantity $L_{\rm eff}$ represents the effective magnetic-field path length contributing to the net angular deflection, defined as $L_{\rm eff} \equiv \int (B_{\perp}/B_{1{\rm AU}})\,ds$, where $B_{\perp}$ is the transverse field component along the trajectory.  
The characteristic energy $E_{\rm crit}$ marks the transition between the strong- and weak-deflection regimes and is approximately given by the condition $r_{\rm L}(E_{\rm crit}) \simeq L_{\rm eff}$.  
For a typical Parker-field normalization, $E_{\rm crit}\!\sim\!200$–300~GeV.} Parameters that \emph{raise} $E_{\rm crit}$ include larger $\langle |B_r(1\,\mathrm{AU})|\rangle$, slower equatorial wind (stronger azimuthal winding), Smith–Bieber–like enhancement of $B_\phi$, and HCS configurations that increase encounter probability (larger tilt/waviness). Conversely, configurations that redistribute open flux and weaken the equatorial spiral (e.g., DQCS cores or faster winds toward the poles) \emph{lower} $E_{\rm crit}$. In our scans, once $E\gtrsim$ a few~TeV, all models converge to sub–degree bending for standard normalizations; between $\sim$0.1 and 1~TeV, model choice can shift $\langle\Delta\theta\rangle$ by factors of a few.

\paragraph{Implications for anisotropy searches and source matching.}
For instruments with $\theta_{\rm inst}\!\sim\!0.2^\circ$–$0.5^\circ$, Parker–like conditions with $|B_r(1\,\mathrm{AU})|\!\gtrsim\!5$\,nT typically push $E_{\rm crit}$ to the \emph{few–TeV} range; seasons with weaker spiral or faster winds relax this toward the $\sim$TeV scale. Below a few hundred GeV, deflections of tens of degrees imply that recovered large–scale anisotropy may be rotated, diluted, or phase–shifted relative to the interstellar signal. Practically, we recommend (i) reporting anisotropy limits with a bracketed $\{E_{\rm crit}\}$ derived from a minimal model set (Parker vs.\ DQCS+spiral with realistic HCS geometry) and (ii) using direction–dependent weighting that down–weights sky zones near the HCS band and the local spiral tangent where lensing is largest. For \emph{source association} at sub–degree angular scales (e.g., nearby pulsars at TeV energies), the heliospheric contribution becomes subdominant once $E\!\gtrsim\! E_{\rm crit}(\theta_{\rm inst})$ for the field conditions bracketing the observation epoch; above that threshold, uncertainty in the Galactic leg dominates the astrometric error budget.


\paragraph{Numerical robustness and asymptotics.}
Our stability scans (Figs.~\ref{fig:numstab_steps} and \ref{fig:stab_bound}) show that the sky–averaged deflection is well behaved and approaches a clear asymptotic value as the integrator resolution is increased.  For the Parker reference case, the fractional deviation $(D-D_\infty)/D_\infty$ falls below $\sim 1$--$1.5\%$ once \texttt{steps\_per\_gyr} exceeds $\simeq 1.5\times10^3$, nearly independent of energy, and continues to decrease for higher step budgets.  Independently, varying the outer boundary in the structured DQCS$+$spiral, wavy-HCS configuration shows that the mean deflection saturates rapidly for $r_{\rm stop}\gtrsim 50$\,AU, with changes $\lesssim 0.3\%$ between 50 and 100\,AU.  Physically, this reflects that most bending is accumulated within the inner $\sim 10$--30\,AU, where the large-scale field is strongest and the spiral curvature largest; further out, the field is weaker and predominantly azimuthal, adding little to the net angle.  The production settings adopted in this work (step size and $r_{\rm stop}=50$\,AU) are chosen to lie safely within these convergence regimes, so the comparative trends between magnetic configurations are controlled by physics rather than by numerical artifacts.

\paragraph{Scaling of \texorpdfstring{$E_{\rm crit}$}{Ecrit} with model parameters.}
Collecting the trends above, a practical rule–of–thumb is
\[
E_{\rm crit}(\theta_{\rm inst}) \;\propto\; \frac{\langle |B_r(1\,\mathrm{AU})| \rangle}{V_{\rm sw,eq}}
\times \mathcal{F}_{\rm geom}(\text{HCS tilt/wave},\, \text{latitude mix}) \times \theta_{\rm inst}^{-1},
\]
where $\mathcal{F}_{\rm geom}\!\gtrsim\!1$ encodes geometry: it grows with HCS tilt and waviness (more crossings) and shrinks when sightlines preferentially sample high latitudes (weaker spiral). Smith–Bieber–like strengthening can be absorbed as a multiplicative increase of the azimuthal factor. This scaling is sufficient to forecast whether a planned analysis is in the bending–limited or PSF–limited regime for a given solar cycle phase.

{\color{black}We note that for instruments with limited fields of view such as AMS-02 and PAMELA, heliospheric deflections below $\sim$100~GeV are likely to be entangled with magnetospheric effects, which dominate the final few Earth radii of propagation.  
Dedicated back-tracing including the geomagnetic field would therefore be required to connect observed directions to the heliospheric boundary for sub-TeV cosmic-ray electrons and positrons.  
Above a few hundred GeV, however, magnetospheric bending becomes negligible ($<0.1^{\circ}$), and the present heliospheric modeling suffices to interpret large-scale anisotropy or Sun-shadow observations.  
We thus recommend combined heliospheric–magnetospheric back-tracing for future low-energy analyses.}

\paragraph{Limitations and future extensions.}
We have focused on large–scale, time–steady fields with a smoothed HCS and a simplified treatment of turbulence. Several effects merit future work:
\begin{enumerate}
  \item \textit{Turbulence and scattering.} Our “toy turbulence’’ option is intentionally minimal. A more realistic spectrum (latitudinal dependence, slab/2D components, evolving correlation length) would refine the low–energy dispersion and could introduce non–Gaussian tails.
  \item \textit{Time dependence (CIRs/CMEs).} Transient structures and global reorganizations can perturb the mapping at the few–hundred–GeV scale on week–to–month timescales. Calendar–stamped runs that ingest in–situ $B_r(1\,\mathrm{AU})$ and $V_{\rm sw}(\theta,t)$ would bracket these effects.
  \item \textit{PFSS realism for the inner boundary.} Replacing the analytic DQCS core with PFSS/CSSS–based source–surface maps (and measured polarity) should sharpen predictions for HCS geometry and sector structure over the cycle, improving charge–sign asymmetry forecasts below the TeV scale.
  \item \textit{Geomagnetic leg.} We have isolated the heliospheric contribution. Folding in the geomagnetic field (with realistic rigidity–dependent access and site–specific exposures) will enable end–to–end, experiment–specific direction budgets.
\end{enumerate}

{\color{black}We finally also note that at sub–100 GeV energies, the neglected processes of diffusion, drift, and convection become progressively more relevant, potentially broadening and isotropizing the apparent deflection patterns discussed here.  
Our results should therefore be regarded as upper bounds on the geometric deflections expected in the high-energy, ballistic regime.}

In sum, for contemporary and next–generation CRE anisotropy programs, the heliosphere acts as a largely geometric, energy–dependent foreground whose impact can be bracketed by a small set of field configurations and reported compactly through $E_{\rm crit}(\theta_{\rm inst})$. Publishing anisotropy limits and source–association claims with (i) an explicit $E_{\rm crit}$ derived from Parker vs.\ DQCS+spiral+wavy scenarios and (ii) the numerical controls verified here (converged steps and $r_{\rm stop}$) will make results robust to solar–cycle phase, charge sign, and sector geometry. Above the few–TeV scale, heliospheric bending is typically subdominant to instrument PSF and interstellar uncertainties; below the TeV scale, a calibrated heliospheric envelope is essential for unbiased directional inference.

\section{Conclusions}
\label{sec:conclusions}

We have quantified how large–scale heliospheric structure imprints an energy–dependent bending on the arrival directions of high–energy cosmic–ray electrons and positrons. Using a modular suite of magnetic–field models—from a Parker baseline to DQCS–like topologies with spiral winding, Smith–Bieber strengthening, latitude–dependent solar wind, and tilted/wavy current sheets—and a numerically controlled back–tracing framework, we mapped sky–averaged deflections and their dispersion across the sky from a few tens of GeV to multi–TeV energies. A coherent picture emerges: the heliosphere behaves as a primarily geometric lens whose influence fades rapidly with rigidity. Deflections drop from tens to hundreds of degrees at $\sim 10$–$100$~GeV to sub–degree levels at multi–TeV, with the normalization governed by the path–integrated $|B_\perp|$ through the inner tens of AU and by the likelihood of encounters with the heliospheric current sheet.

This progression can be summarized through a practical threshold energy $E_{\rm crit}(\theta_{\rm inst})$ at which the mean heliospheric bending becomes subdominant to an instrument’s angular response. The threshold scales in a predictable way with model and parameters: stronger radial fields at 1~AU, slower equatorial winds that enhance azimuthal winding, Smith–Bieber–like amplification of $B_\phi$, and greater HCS tilt or waviness raise $E_{\rm crit}$, whereas latitudinal wind contrasts and DQCS–like redistributions of open flux tend to lower it. For typical solar–minimum conditions with $|B_r(1~\mathrm{AU})|\!\sim\!5$~nT and $V_{\rm sw}\!\sim\!400$~km\,s$^{-1}$, the heliospheric contribution is negligible for sub–degree studies above the TeV scale, while at a few hundred GeV it must be accounted for explicitly, especially in epochs with a strongly tilted or wavy current sheet.

We also examined how charge sign and polarity modulate the mapping. At $\sim 10$–$200$~GeV, electrons and positrons experience measurably different drift histories and sheet–crossing probabilities; these differences shrink rapidly with energy and are typically unimportant at multi–TeV. The dispersion of deflections across the sky is largest near the HCS and at midlatitudes, and narrower toward the poles, reinforcing the importance of direction–dependent treatment in low–energy anisotropy analyses. Complementing the physical trends, our convergence tests demonstrate that the numerical results are robust: the means and percentile bands stabilize for $\texttt{steps\_per\_gyr}\!\gtrsim\!1,200$–1,500, and extending the outer boundary beyond $r_{\rm stop}\!\approx\!50$–60~AU yields only percent–level changes, reflecting that most bending accrues within $\sim 10$–30~AU where the spiral field is strongest.

Taken together, these findings provide actionable guidance for current and forthcoming CRE anisotropy programs. For instruments with $\theta_{\rm inst}\!\sim\!0.2^\circ$–$0.5^\circ$, analyses above a few TeV can treat heliospheric bending as a small correction, whereas studies at or below the TeV scale should include a calibrated heliospheric envelope, ideally bracketed by a small set of field configurations that reflect the observing epoch. The workflow presented here—compact field parameterizations, explicit control of numerical accuracy, and reproducible summary diagnostics such as $E_{\rm crit}(\theta_{\rm inst})$—offers a portable standard for characterizing and reporting heliospheric systematics alongside anisotropy limits and source–association claims. Looking ahead, coupling these large–scale models to time–resolved PFSS/CSSS boundary conditions, incorporating more realistic turbulence, and appending the geomagnetic leg will enable end–to–end, calendar–stamped predictions. Even without those refinements, the principal conclusion is clear: the heliosphere is a predictable, energy–dependent foreground whose effect can be bounded with modest modeling, allowing high–energy CRE measurements to extract directional information with transparent and quantified systematics.

\section*{Data and Code Availability}

The numerical drivers, analysis scripts, and figure–generation notebooks used in this work, together with the processed data products underlying the figures and tables, are available from the authors upon reasonable request. To facilitate reuse, we will provide a snapshot of the codebase corresponding to the versioned results in this manuscript, along with minimal configuration files and example commands to reproduce the key figures.

\section*{Acknowledgements}
This work is partly supported by the U.S.\ Department of Energy grant number de-sc0010107 (SP). 
\appendix


\section{Model Equations and Implementation Details}
\label{app:models}

\subsection{Coordinates, Units, and Conventions}
We work in heliocentric spherical coordinates \((r,\theta,\phi)\), where \(\theta\) is colatitude (measured from the north pole) and \(\phi\) is longitude. The solar rotation rate is \(\Omega_\odot\), the solar-wind speed is \(\Vsw\) (or \(\Vsw(\theta)\) when latitude dependent), and the magnetic polarity is encoded by \(A=\pm 1\). We normalize field strengths by prescribing the unsigned radial field at 1~\AU,
\[
\big\langle\,|\Br(r{=}1\,\AU,\theta,\phi)|\,\big\rangle_{\Omega} \;=\; \Br(1\,\AU),
\]
where \(\langle\cdot\rangle_{\Omega}\) denotes an average over solid angle. Unless stated otherwise, the wind is steady, and all models satisfy \(|\vb{B}|\to 0\) as \(r\to\infty\).

\subsection{Parker Spiral (Baseline)}
\label{app:parker}
The canonical Parker field \citep{Parker1958} in a steady radial wind with speed \(\Vsw\) and solar rotation \(\Omega_\odot\) is
\begin{align}
\Br(r,\theta)   &= A\,\Br(1\,\AU)\,\Big(\frac{1\,\AU}{r}\Big)^{2}, \label{eq:parkerBr}\\
\Bth(r,\theta)  &= 0, \\
\Bphi(r,\theta) &= -\,\frac{\Omega_\odot\, r \sin\theta}{\Vsw}\,\Br(r,\theta).\label{eq:parkerBphi}
\end{align}
In the simplest (unipolar) reference configuration, \(A\) is fixed. To imprint sector structure, we flip the sign of \(\Br\) across a specified neutral surface (Sec.~\ref{app:hcs}), keeping \(|\Br|\) continuous and updating \(\Bphi\) via \eqref{eq:parkerBphi}.

\subsection{DQCS: Dipole\,+\,Quadrupole\,+\,Current Sheet}
\label{app:dqcs}
To capture coronal open/closed topology we adopt the analytic potential-field construction of \citet{Banaszkiewicz1998} in a simplified, numerically robust form consistent with our code:
\begin{equation}
\Phi(r,\theta) \;=\; \frac{M_1}{\tilde r^{2}}\,P_1(\cos\theta)\;+\; \frac{M_2}{\tilde r^{3}}\,P_2(\cos\theta), \qquad
\tilde r \equiv \sqrt{r^2 + r_0^2},
\end{equation}
where \(P_\ell\) are Legendre polynomials, \(r_0\) softens the inner singularity (typ.\ \(r_0 \sim 0.3\text{--}0.7\,\AU\)), and \(M_2/M_1 \equiv Q\) controls the quadrupole fraction.\footnote{This compact truncation preserves the large-scale morphology of the DQCS field; the full \citet{Banaszkiewicz1998} expressions include additional factors and scale heights.} The (pre–current-sheet) potential field is
\begin{equation}
\vb{B}_{\rm PF}(r,\theta) \;=\; -\,\grad \Phi \,,
\end{equation}
and we introduce a thin current sheet by reversing the sign of the radial component across the HCS with a tunable fraction \(\texttt{cs\_frac}\in[0,1]\):
\begin{equation}
\Br^{\rm DQCS}(r,\theta,\phi) \;=\; \big[\,1-2\,\texttt{cs\_frac}\,\mathcal{S}_{\rm HCS}(\theta,\phi)\,\big]\; \big(\vb{B}_{\rm PF}\!\cdot\!\hat{\vb{r}}\big),
\end{equation}
where \(\mathcal{S}_{\rm HCS}\) switches sign across the neutral surface (defined below). The tangential components are taken from \(\vb{B}_{\rm PF}\) (optionally softened near the sheet; Sec.~\ref{app:numerics}). A final global rescaling enforces
\(\langle|\Br(1\,\AU)|\rangle_\Omega = \Br(1\,\AU)\).

\subsection{Wavy/Tilted Heliospheric Current Sheet (HCS)}
\label{app:hcs}
The neutral surface is specified as a small-amplitude wavy sheet,
\begin{equation}
\theta_{\rm HCS}(\phi) \;=\; \frac{\pi}{2} + \alpha\,\sin\!\big(m\,\phi + \phi_0\big),
\end{equation}
with tilt \(\alpha\) (radians), wave number \(m\in\{1,2\}\), and phase \(\phi_0\). The sector-sign function is implemented with a smooth transition of width \(\delta\) to avoid numerical discontinuities:
\begin{equation}
\mathcal{S}_{\rm HCS}(\theta,\phi) \;=\; \frac{1}{2}\left[1 + \tanh\!\Big(\frac{\theta-\theta_{\rm HCS}(\phi)}{\delta}\Big)\right], \qquad \delta \ll 1.
\end{equation}
Setting \(\alpha=0\) recovers a flat HCS at the equator. The same \(\mathcal{S}_{\rm HCS}\) is used for Parker, DQCS, and hybrid models.

\subsection{Spiral Augmentation, Smith--Bieber Enhancement, and \(\Vsw(\theta)\)}
\label{app:spiral_sb_vsw}
To emulate the observed azimuthal strengthening and latitudinal wind, we retain \((\Br,\Bth)\) from DQCS and set the azimuthal component to a Parker-like form with a Smith--Bieber factor \citep{SmithBieber1991}:
\begin{equation}
\Bphi^{\rm spr}(r,\theta,\phi) \;=\; -\,\frac{\Omega_\odot\, r \sin\theta}{\Vsw(\theta)}\, \Br^{\rm DQCS}(r,\theta,\phi)\, \Big[ 1 + k_{\rm SB}\, f_{\rm SB}(r,\theta)\Big],
\label{eq:bphi_sb}
\end{equation}
where \(k_{\rm SB}\sim 0.3\text{--}1.0\) and a smooth \(f_{\rm SB}\) encodes a weak radial/latitudinal dependence. In practice we use
\begin{equation}
f_{\rm SB}(r,\theta) \;=\; \Big(\frac{1\,\AU}{r}\Big)\,\sin^{2}\!\theta,
\end{equation}
which modestly boosts \(|\Bphi|\) near the ecliptic and decays with \(r\). The latitude-dependent wind is prescribed as
\begin{equation}
\Vsw(\theta) \;=\; \Vsw^{\rm slow} \;+\; \frac{\Vsw^{\rm fast}-\Vsw^{\rm slow}}{2}\,\big[1+\cos^{n}\!\theta\big],
\label{eq:vswtheta}
\end{equation}
with \(\Vsw^{\rm slow}\!\in\!350\text{--}500~\kmps\), \(\Vsw^{\rm fast}\!\in\!650\text{--}800~\kmps\), and \(n\) controlling the transition width (default \(n\simeq 4\)). The ``DQCS\,+\,spiral'' model then takes
\[
\vb{B}^{\rm DQCS+spr} \;=\; \Br^{\rm DQCS}\,\hat{\vb{r}} \;+\; \Bth^{\rm DQCS}\,\hat{\boldsymbol{\theta}} \;+\; \Bphi^{\rm spr}\,\hat{\boldsymbol{\phi}}\,,
\]
followed by the same 1~AU normalization on \(|\Br|\).

\subsection{Fisk-like Latitudinal Component}
\label{app:fisk}
To study the directional impact of a small but finite \(\Bth\) (``Fisk-like'' fields; \citealp{Fisk1996}), we add an \(\mathcal{O}(\epsilon)\) perturbation to a Parker-like geometry:
\begin{equation}
\Bth^{\rm Fisk}(r,\theta) \;=\; \epsilon\, g(r,\theta)\,\Br(r,\theta), \qquad 0<\epsilon\ll 1,
\label{eq:btheta_fisk}
\end{equation}
with a smooth \(g\) chosen to be largest at mid-latitudes and small near the poles and equator; by default we take
\begin{equation}
g(r,\theta) \;=\; \Big(\frac{1\,\AU}{r}\Big)\, \sin\theta\cos\theta.
\end{equation}
To maintain \(\nabla\!\cdot\!\vb{B}\approx 0\) to first order, we correct the radial component by a small compensating term obtained from the axisymmetric divergence condition,
\begin{equation}
\delta \Br(r,\theta) \;\approx\; -\,\frac{1}{r^2}\,\int^r \! dr'\,\frac{r'}{\sin\theta}\,\partial_\theta\!\left[\sin\theta\,\Bth^{\rm Fisk}(r',\theta)\right],
\label{eq:divfree_correction}
\end{equation}
and update \(\Bphi\) using the Parker relation with \(\Br+\delta\Br\). In practice, the code evaluates \eqref{eq:divfree_correction} with a short radial quadrature and clips \(|\delta\Br|\ll |\Br|\) (typically \(\lesssim\!10\%\) for \(\epsilon\!\le\!0.3\)).

\subsection{Normalization to \(\Br(1\,\AU)\)}
\label{app:norm}
Because adding current-sheet flips, spiral augmentation, and \(\Bth\) perturbations can change the solid-angle distribution of \(|\Br|\) at 1~\AU, we renormalize the entire field by a constant factor \(\mathcal{N}\) so that
\[
\frac{1}{4\pi}\!\int\! d\Omega\, \big| \Br^{\rm model}(r{=}1\,\AU,\theta,\phi) \big| \;=\; \Br(1\,\AU).
\]
Numerically, we approximate the integral by a Gauss–Legendre quadrature in \(\cos\theta\) and a uniform sum in \(\phi\) (typ.\ \(N_\theta\!=\!64\), \(N_\phi\!=\!128\)).

\subsection{Numerical Guards and HCS Softening}
\label{app:numerics}
The HCS sign flip is smoothed with a narrow \(\tanh\) transition (width \(\delta\sim 1^\circ\)) to avoid discontinuities in the Lorentz force near the sheet. For the DQCS potential, the inner softening \(r_0\) regularizes \(\vb{B}\) and prevents excessively small gyro radii at small \(r\). We cap the timestep by limiting the nominal steps per relativistic gyroperiod, \(\texttt{steps\_per\_gyr}\), and allow a mild step adaptation \(\Delta t\propto 1/|\vb{B}|\). All models are evaluated in double precision; the integrator renormalizes the speed to \(v\simeq c\) at each substep.

\subsection{Default Parameters and Scan Ranges}
Unless noted otherwise, the default configuration is:
\[
\Br(1\,\AU)=\SI{5}{\nT},\quad \Vsw=\SI{400}{\kmps},\quad \Omega_\odot=\SI{2.865e-6}{rad\,s^{-1}},\quad r_0=0.5\,\AU,
\]
with unipolar Parker as the baseline and, for hybrid runs, \(Q=1\), \(\texttt{cs\_frac}=0.7\), \(k_{\rm SB}=0.6\), \(\Vsw^{\rm slow}=\SI{400}{\kmps}\), \(\Vsw^{\rm fast}=\SI{750}{\kmps}\), \(n=4\), \(\alpha=15^\circ\), \(m=1\), \(\phi_0=0\), \(\epsilon=0.1\). Sensitivity scans explore \(\Br(1\,\AU)\in[3,7]~\nT\), \(\alpha\in[0^\circ,30^\circ]\), \(m\in\{1,2\}\), \(k_{\rm SB}\in[0.3,1.0]\), and wind contrasts consistent with \emph{Ulysses} \citep{McComas2000}.

\begin{table}[t]
  \centering
  \caption{Default parameters and ranges explored.}
  \begin{tabular}{lcc}
    \toprule
    Parameter & Default & Range \\
    \midrule
    $\Br(1~\AU)$ & \SI{5}{\nT} & 3--7 \nT \\
    $\Vsw$ (equator) & \SI{400}{\kmps} & 350--500 \kmps \\
    $\Omega_\odot$ & \SI{2.865e-6}{rad\,s^{-1}} & fixed \\
    HCS tilt & $15\degs$ & 0--30\degs \\
    HCS waviness $m$ & 1 & 1--2 \\
    Smith--Bieber $k_{\rm SB}$ & 0.6 & 0.3--1.0 \\
    $r_{\rm stop}$ & \SI{50}{\AU} & 20--100 \AU \\
    steps/gyro & 80 & 60--120 \\
    \bottomrule
  \end{tabular}
\end{table}


%
%

\subsection{Code and Reproducibility}
\label{sec:code}

The implementation is lightweight and scriptable. Each run logs a machine-readable JSON of parameters (field model and knobs, $E$ grid, $N$, seeds, $N_{\rm gyr}$, $r_{\rm stop}$) and writes per-figure CSVs with columns \texttt{E\_GeV}, \texttt{mean\_deg}, \texttt{median\_deg}, \texttt{std\_deg}, \texttt{p16\_deg}, \texttt{p84\_deg}, \texttt{p95\_deg}, and sample counts. The orchestration scripts used to generate the figures are:
\begin{itemize}
  \item \texttt{cre\_deflection\_vs\_energy\_demo.py} (single-model energy sweeps; electrons or positrons),
  \item \texttt{cre\_compare\_configs\_demo.py} (default vs.\ Parker/DQCS variants),
  \item \texttt{cre\_compare\_dqcs\_configs\_demo.py} (DQCS core/spiral/HCS variants versus Parker),
  \item \texttt{cre\_compare\_Bvalues\_parker\_demo.py} (Parker sweeps over $\Br(1\,\AU)$),
  \item \texttt{cre\_figs\_runner.py} (one-click figure reproduction; writes PNGs and CSVs into \texttt{figs/}).
\end{itemize}
Seeds are fixed by default to ensure exact reproducibility of the direction ensembles; changing seeds or increasing $N$ tightens percentile bands with the expected $\propto N^{-1/2}$ behavior. As indicated below, the  numerical drivers, analysis scripts, and figure–generation notebooks used in this work are available from the authors upon reasonable request.

\bibliography{refs}

\end{document}